\documentclass{article}

\usepackage{arxiv}

\usepackage[utf8]{inputenc} 
\usepackage[T1]{fontenc}    
\usepackage{hyperref}       
\usepackage{url}            
\usepackage{booktabs}       
\usepackage{amsfonts}       
\usepackage{nicefrac}       
\usepackage{microtype}      
\usepackage{lipsum}		
\usepackage{graphicx}
\usepackage{natbib}
\usepackage{doi}

\usepackage[english]{babel}

\usepackage{lineno}
\usepackage{float}
\usepackage{booktabs}
\usepackage{amsfonts}
\usepackage{amsmath}
\usepackage{amsthm}
\usepackage{xcolor}
\usepackage{amssymb}
\usepackage{graphicx}
\usepackage{algorithm}
\usepackage{algpseudocode}
\usepackage[normalem]{ulem}
\hypersetup{
    colorlinks = true,
    linkcolor = blue,
    anchorcolor = blue,
    citecolor = blue,
    filecolor = blue,
    urlcolor = blue
    }

\usepackage{tikz}
\DeclareRobustCommand\full  {\tikz[baseline=-0.6ex]\draw[thick] (0,0)--(0.4,0);}
\DeclareRobustCommand\dotted{\tikz[baseline=-0.6ex]\draw[thick,dotted] (0,0.0)--(0.44,0);}

\theoremstyle{definition}
\newtheorem{definition}{Definition}[section]
\newtheorem{prop}{Proposition}[section]

\newenvironment{myproof}[1][\proofname]{%
  \begin{proof}[#1]$ $\par\nobreak\ignorespaces
}{%
  \end{proof}
}

\usepackage{caption}
\usepackage{subcaption}
\newcommand{\blue}[1]{\textcolor{blue}{#1}}
\newcommand{\red}[1]{\textcolor{red}{#1}}

\newcommand{\gray}[1]{\textcolor{gray}{#1}}
\newcommand{\lightgray}[1]{\textcolor{lightgray}{#1}}

\title{Domain Selection for Gaussian Process Data: An application to
electrocardiogram signals}

\author{ \href{}{\hspace{1mm}Nicolás Hernández} \\
	Department of Statistical Science\\
	University College London\\
	London, UK \\
	\texttt{n.hernandez@ucl.ac.uk} \\
	\And
	\href{}{\hspace{1mm}Gabriel Martos} \\
	Departmento de Matemáticas y Estadística\\
	Universidad Torcuato Di Tella\\
    Buenos Aires, Argentina \\
	\texttt{gmartos@utdt.edu} \\
}

\renewcommand{\shorttitle}{Domain Selection for Gaussian Process Data}
\hypersetup{
pdftitle={Domain Selection for Gaussian Process Data},
pdfsubject={},
pdfauthor={N. Hernández, G. Martos},
pdfkeywords={Domain selection; Gaussian processes; Kullback-Leibler divergence; intervals of local maximum divergence; electrocardiogram signals.},
}

\begin{document}
\maketitle

\begin{abstract}
Gaussian Processes and the Kullback-Leibler divergence have been deeply studied in Statistics and Machine Learning. This paper marries these two concepts and introduce the local Kullback-Leibler divergence to learn about intervals where two Gaussian Processes differ the most. We address subtleties entailed in the estimation of local divergences and the corresponding interval of local maximum divergence as well. The estimation performance and the numerical efficiency of the proposed method are showcased via a Monte Carlo simulation study. In a medical research context, we assess the potential of the devised tools in the analysis of electrocardiogram signals.
\end{abstract}


\section{Introduction}\label{Intro}

Everyday millions of complex data patterns flow around the world at unprecedented speed, leading to an explosion on the demand for modelling random process data, such as time series and functional data. Electrocardiogram signals (ECG) are an example of such high-dimensional data, that is usually structured in the form of curves almost continuously recorded over a grid of discrete time points. From a medical point of view, when a patient is admitted in the emergency room during a cardiac arrest, one of the few pieces of information available to make a diagnosis is an ECG. It is therefore an extremely useful tool for immediate decision making and it can also be helpful in determining the causes and the gravity of a cardiac pathology \cite{mullainathan2022solving}. Nevertheless, the analysis of ECG data faces important challenges in practice, in particular due to its high dimensionality. Therefore, the study of local features of ECG signals play a key role in Medicine \cite{wang2013human,rodriguez2015feature, mullainathan2022solving} for at least two important reasons: 
\begin{enumerate}
    \item[i)] Diagnosis: Identifying time intervals during the cardiac cycle where the ECG signal present atypical patterns is crucial in order to improve the early diagnosis of different cardiac diseases, increasing the patient survival probability during a cardiac episode.
    \item[ii)] Causes and effects: 
    Learning time intervals with an anomalous pattern during the cardiac cycle will help to understand the origins and the consequences of different heart diseases.
\end{enumerate}
Taking into account (i) and (ii) above, the main goal of this paper is to learn from ECG data an \textit{interval with a certain length} where the signals corresponding to disease and healthy subjects differ the most. Throughout this paper, we refer to this problem as that of \textit{domain selection} for ECG signals, but the method devised in this paper also apply to other random processes data in a broad sense. We model ECG signals using Gaussian Processes (GPs), a versatile and flexible tool for modelling complex signal patterns  \cite{perez2013gaussian},  and introduce the \emph{local} Kullback-Leibler (KL) divergence \cite{kullback1951information} as we formally discussed in Section~\ref{DomainSelection}, so to learn intervals of maximum local divergence between GPs. Recent contributions in the related literature explore \textit{domain selection} methods--a.k.a. variable selection in functional data   \cite{berrendero2016variable,baillo2011classification,pini2017interval}-- to achieve accurate prediction for functional data classification methods and to assess local differences between functional means in a two sample problem. Also in the context of statistics in medicine, the authors in  \cite{martos2018} propose a Mann-Whitney type of statistic for functional data to learn about intervals at which two processes differ the most, based on aspects related with symmetry. Our approach differs from the ones mentioned above in at least two important ways: (1) Here the ultimate goal is neither to classify nor to test hypothesis with random processes data, but rather to \textbf{learn the interval with a given length where two random processes}, which corresponds to groups of ECG signals, \textbf{differ the most}; (2) Our approach relies on GP and the KL divergence, whereas the aforementioned methodologies have mainly been designed in the context of functional data. As a byproduct, we also contribute on the following points:
\begin{itemize}
\item Optimisation: We introduce the interval of local maximum divergence through a set function optimisation problem. Therefore, the estimation methods proposed in the paper contributes to the literature on set function optimisation. 
\item Classification: When the analysis of ECG signals also entails the discrimination between groups (i.e. healthy vs disease), our method could benefit other standard functional classifiers if they are applied on a 
small interval where the two processes 
differ the most, rather than treating the entire signal domain equally. 
\item Storage efficiency: If only a subset of the entire domain is found to be relevant in order to assess differences between healthy subjects and disease patients, then this suggests the potential benefit for collecting and saving only a smaller subset of ECG signals. 

\item Miscellany: In the paper we also establish conditions for the existence of an interval of local maximum divergence, consider subtleties entailed in the estimation of local KL divergences and the interval of local maximum divergence; and also discuss variants and extensions around the proposed methodology.
\end{itemize}

The remind of the paper is organised as follows: In Section~\ref{DomainSelection} we introduce the local KL divergence for GP and the interval of local maximum divergence, and also discuss suitable corresponding estimation methods. In Section~\ref{S:Experiments} we present Monte Carlo evidence to assess the consistency of our estimator, while in Section~\ref{s:realdata} we illustrate the method with an electrocardiogram signals data application. Finally in Section~\ref{S:discusion} we discuss the results and conclude our work.

\section{Materials and Methods}\label{DomainSelection}
The goal in this section is to introduce the probabilistic framework to assess local differences between GPs. To this end, let $X\sim N(\mu_X,\sigma_X)$ and $Y\sim N(\mu_Y,\sigma_Y)$ be two normally distributed independent random variables, where $\mu_\ell$ and $\sigma_\ell$ for $\ell \in \{X,Y\}$ denotes the corresponding mean and variance parameters; then the KL divergence can be written in a closed from as follows:
\begin{equation}\label{KL-gaussian}
\text{KL}(X||Y) = \frac{1}{2} \left(\frac{\sigma^2_X}{\sigma^2_Y}-1 +\frac{(\mu_X- \mu_Y)^2}{\sigma^2_Y} 
+ \ln\left( \frac{\sigma^2_Y}{\sigma^2_X}\right) \right) .
\end{equation}
The divergence in Eq.~\eqref{KL-gaussian} is a functional that quantifies the dissimilarity in the distribution of two Gaussian random variables $X$ and $Y$; and more importantly Eq.~\eqref{KL-gaussian} is easy and computationally cheap to evaluate when $X$ and $Y$ are multivariate normal random vectors. Next we discuss how to extrapolate Eq.~\eqref{KL-gaussian} to compute intervals of local maximum divergence for GPs.

\subsection{Local Kullback-Leibler divergence for Gaussian Processes}\label{ss:KL}
Let $X(t)\sim GP(\mu_X(t),\sigma_X(t,s))$ be a GP with mean function $\mu_X(t) = E\{X(t)\}$ and variance function $\sigma_X(t,s) = E\{(X(t)-\mu_X(t))(X(s)-\mu_X(s))\}$, indexed on the compact set $T\subset \mathbb{R}$;  we address the following problem on the comparison of two GP:

\begin{quote}\textbf{Learning Problem}. \textit{Given data drawn from GPs $X(t)\sim GP(\mu_X(t),\sigma_X(t,s))$ and $Y(t)\sim GP(\mu_Y(t),\sigma_Y(t,s))$, indexed on the same compact domain $T\subset \mathbb{R}$, learn the interval with a given certain length where they statistically differ the most.}
\end{quote}

\begin{figure}[h!]
\centering 
\includegraphics[width=0.9\textwidth]{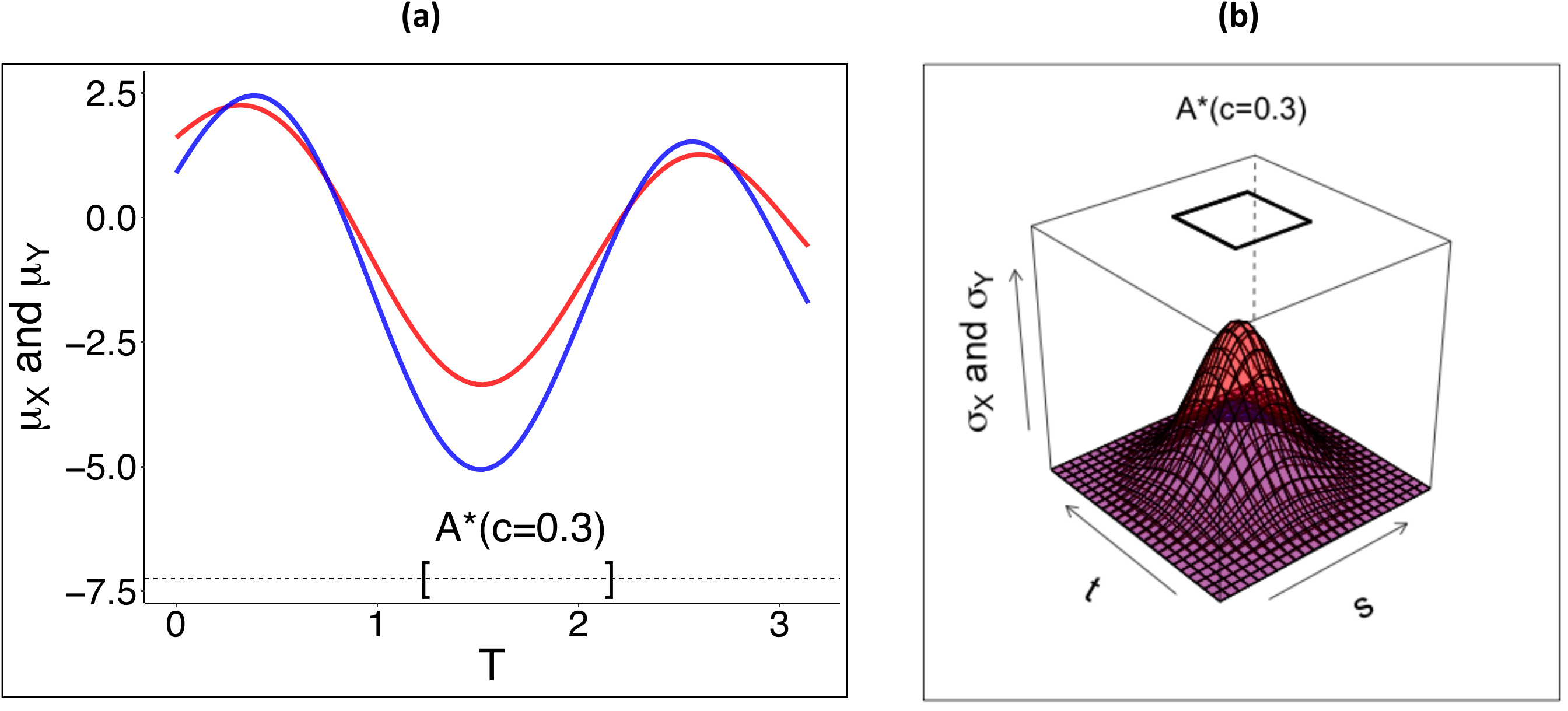}
\caption{\textbf{(a)} Mean functions $\mu_{X}(t)$ and $\mu_{Y}(t)$. \textbf{(b)}  Variance functions $\sigma_{X}(t,s)$ and $\sigma_{Y}(t,s)$. The corresponding interval of local maximum divergence for $c=0.3$ depicted in brackets in \textbf{(a)} and with a black square in \textbf{(b)}. \label{fig:1}}
\end{figure}

Since the mean and variance functions completely determines the law of $X(t)$ and $Y(t)$; hereafter we assume that there exists 
a compact subset $A\subset T$ with $\lambda(A)>0$, where  $\lambda(A)$ is the Lebesgue measure of the set $A$, and a positive constant $\nu$, such that one of the following scenarios of local differences holds:
\begin{enumerate}
\item[(A)] For every $t\in A$: $|\mu_X(t)- \mu_Y(t)|>\nu$, while for all $t^\prime\notin A$ it holds that $|\mu_X(t^\prime)- \mu_Y(t^\prime)|\leq \nu$.
\item[(B)] For every $(t,s)\in A\times A$: $|\sigma_X(t,s) -\sigma_Y(t,s)|>\nu$, while for all $(t^\prime,s^\prime)\notin A\times A$ it holds that $|\sigma_X(t^\prime,s^\prime) -\sigma_Y(t^\prime,s^\prime)|\leq \nu$.
\item[(C)] Local mean and variance differences corresponds to scenario (A) and (B) simultaneously, possibly on different subsets $A_\mu$ and $A_\sigma$ and for different thresholds constant $\nu_\mu$ and $\nu_\sigma$.
\end{enumerate}
In Figure~\ref{fig:1}--(a) we illustrate these scenario A: The coloured lines represents the mean functions of two GP; notice that over a relatively small time interval--around the time point $t=1.5$-- the difference $|\mu_X(t)- \mu_Y(t)|$ exceeds a certain threshold $\nu_\mu$. In Figure~\ref{fig:1}-(b) we depict scenario B: The coloured surfaces represents the covariance functions of two GP; notice that over a relatively small time interval--highlighted with a black square on the top-- the difference $|\sigma_X(t,s)- \sigma_Y(t,s)|$ is above a certain threshold $\nu_\sigma$. The scenario C corresponds to Figure~\ref{fig:1}--(a) and (b) simultaneously. 

The goal of this paper is to develop a local KL divergence measure and related estimation methods to learn about intervals or regions, such as $A^*(c)$ in Figure~\ref{fig:1}-(a) and (b), where two GP's differ the most (i.e. where the functional parameters are the most dissimilar). To this end, it will be technically convenient and computationally efficient to use a finite-dimensional (or discrete) representation of GP data. Without loss of generality, we assume that data is recorded and stored over the same discrete and equally spaced grid of points $\mathcal{T}=(t_1,\dots,t_p) \subset T$, being $p\gg 0$ (other more general cases are easy to tackle as we discuss in Section~\ref{S:Experiments}). In this setting, GP data corresponds to realisations of $p$-variate Gaussian random vectors $\textbf{X}_\mathcal{T} \equiv (X(t_1),\dots,X(t_p))\sim N_p(\boldsymbol{\mu}_{\mathcal{T},X},\boldsymbol{\Sigma}_{\mathcal{T},X})$ and $\textbf{Y}_\mathcal{T} \equiv (Y(t_1),\dots,Y(t_p))\sim N_p(\boldsymbol{\mu}_{\mathcal{T},Y},\boldsymbol{\Sigma}_{\mathcal{T},Y})$ where $\boldsymbol{\mu}_{\mathcal{T},X} = (\mu_X(t_1),\dots,\mu_X(t_p))$ and $\boldsymbol{\mu}_{\mathcal{T},Y} = (\mu_Y(t_1),\dots,\mu_Y(t_p))$ are the corresponding means, and $\boldsymbol{\Sigma}_{\mathcal{T},X} \in \mathbb{R}^{p\times p}$ and $\boldsymbol{\Sigma}_{\mathcal{T},Y} \in \mathbb{R}^{p\times p}$ the corresponding $p\times p$ variance matrices (i.e. $[\boldsymbol{\Sigma}_{\mathcal{T},X}]_{ij}= \sigma_X(t_i,t_j)$ and $[\boldsymbol{\Sigma}_{\mathcal{T},Y}]_{ij}= \sigma_Y(t_i,t_j)$ for $i=1,\dots,p$ and  $j=1,\dots,p$).\\

Under this high dimensional GP's representation, the KL divergence  \cite{kullback1951information} is a natural metric to assess differences in distribution between two GP's. The KL divergence between GP $X$ and $Y$ over the grid $\mathcal{T}$ is computed as follows  \cite{pardo2018statistical}:

\begin{equation}\label{Eq:KL}
\text{KL}_\mathcal{T}(X||Y) \equiv {\frac {1}{2}}\left(\operatorname{tr} \left(\boldsymbol \Sigma_{\mathcal{T},Y}^{-1}\boldsymbol\Sigma_{\mathcal{T},X} -\textbf{I}_p \right)+\Delta_\mathcal{T}^{\mathsf{T}}\boldsymbol\Sigma_{\mathcal{T},Y}^{-1}\Delta_\mathcal{T}+\ln \left({\frac {\det \boldsymbol\Sigma_{\mathcal{T},Y}}{\det \boldsymbol\Sigma_{\mathcal{T},X}}}\right)\right),
\end{equation}
where $\Delta_\mathcal{T} = \left(\boldsymbol\mu_{\mathcal{T},Y}-\boldsymbol\mu_{\mathcal{T},X}\right)$, $\operatorname{tr}(\boldsymbol \Sigma)$ denote the trace of $\boldsymbol \Sigma$ and $\operatorname{det}(\boldsymbol \Sigma)$ the determinant of $\boldsymbol \Sigma$. Some comments on the KL divergence are in order: (i) The expression in Eq.~\eqref{Eq:KL} is a generalisation of the corresponding univariate KL divergence in Eq.~\eqref{KL-gaussian}. (ii) The KL divergence is not symmetric, nevertheless the symmetrisation is straightforward: consider for instance $\text{KL}_\mathcal{T}(X||Y)/2+\text{KL}_\mathcal{T}(Y||X)/2$. 
(iii) Interestingly, the KL divergence considers simultaneously differences in mean and variance, i.e. both mean and variance appear together in Equation~\eqref{Eq:KL}. (iv) In the case of GP with the same covariance function, then $2\text{KL}_\mathcal{T}(X||Y) = \Delta_\mathcal{T}^{\mathsf{T}}\boldsymbol\Sigma_{\mathcal{T},Y}^{-1}\Delta_\mathcal{T}$ corresponds to the squared Mahalanobis distance  \cite{mclachlan1999mahalanobis} between the two GP.

Our goal is to learn subsets of $T$ where the two processes differ the most, therefore the KL divergence is a suitable ``statistical distance" to assess local differences between GP. For any subset $\mathcal{A}\subseteq  \mathcal{T}$, we  define the local-KL divergence as follows:
\begin{equation}\label{Eq:KL_local}
\text{KL}_\mathcal{A}(X||Y) \equiv {\frac {1}{2}}\left(\operatorname{tr} \left(\boldsymbol \Sigma_{\mathcal{A},Y}^{-1}\boldsymbol\Sigma_{\mathcal{A},X} -\textbf{I}_{|\mathcal{A}|} \right)+    \Delta_\mathcal{A}^{\mathsf{T}}\boldsymbol\Sigma_{\mathcal{A},Y}^{-1}\Delta_\mathcal{A}+\ln \left({\frac {\det \boldsymbol\Sigma_{\mathcal{A},Y}}{\det \boldsymbol\Sigma_{\mathcal{A},X}}}\right)\right),
\end{equation}
where $\Delta_\mathcal{A} = \left(\boldsymbol\mu_{\mathcal{A},Y}-\boldsymbol \mu_{\mathcal{A},X}\right)$ and $\{\boldsymbol\mu_{\mathcal{A}},\boldsymbol\Sigma_{\mathcal{A}}\}$ denotes suitable partitions of $\{\boldsymbol\mu_{\mathcal{T}},\boldsymbol\Sigma_{\mathcal{T}}\}$ in correspondence with the subset $\mathcal{A}\subseteq \mathcal{T}$.  Local KL divergences for GP's data also have a number of interesting properties which we summarise next.

\begin{prop}\label{prop1}
The $\text{KL}_{\mathcal{A}}(X||Y)$ is a set function that satisfies the following properties:
\begin{enumerate}
\item[(a)] \textbf{Non-negative:} For fixed GPs $X$ and $Y$, it holds: $\text{KL}_\mathcal{A}(X||Y):\mathcal{P}_\mathcal{T}\to \mathbb{R}^+_0$ where $\mathcal{P_T}$ is the power set of $\mathcal{T}$; and $\text{KL}_\mathcal{A}(X||Y)=0$ if and only if $\mu_X(t)=\mu_Y(t)$ for all $t\in \mathcal{A}$ and $\sigma_X(t,s)=\sigma_Y(t,s)$ for all $(t,s)\in \mathcal{A}\times \mathcal{A}$.
    \item[(b)] The local KL divergence is \textbf{upper bounded} (i.e. $\text{KL}_\mathcal{T}(X||Y) < \infty)$ and a \textbf{monotone} set  function (i.e. for $\mathcal{A}' \subseteq \mathcal{A}$ it holds: $\text{KL}_{\mathcal{A}'}(X||Y)\leq \text{KL}_\mathcal{A}(X||Y)$), under suitable conditions on the mean and variance functions.
    \item[(c)] The local divergence $\text{KL}_\mathcal{A}(X||Y) $ is a \textbf{continuous} set function in $\mathcal{C}_\mathcal{T}$, the collection of all contiguous subsets from the ground set $\mathcal{T}$.
    \end{enumerate}    
    \end{prop}

In the Appendix we give formal proofs on previous assertions. Since $\text{KL}_\mathcal{A}(X||Y)$ accounts simultaneously for differences in mean and variance, we propose this particular metric to define the most \textit{discriminating subset of points} in $\mathcal{T}$ as follows: 

\begin{definition}
Let $|\cdot|$ be the counting measure on $\mathcal{P}_\mathcal{T}$, we define the subset $\widetilde{\mathcal{A}}^*(\tilde{c})\subset \mathcal{T}$, for any $1\leq \tilde{c}\leq |\mathcal{T}|$, as the \textit{variable selection} subset that solves the following set function optimisation problem:
\begin{equation}\label{optim1}
\max_{\mathcal{A} \subset \mathcal{T}} \text{KL}_\mathcal{A}(X||Y), \text{ s.t. } |\mathcal{A}| \leq \tilde{c}.
\end{equation}
\end{definition}
Interestingly, this definition resembles some developments in the context of variable selection for functional data classification \cite{berrendero2016variable}, where the authors propose a maxima--hunting approach to search for isolated time points in the domain corresponding to random processes $X(t)$ and $Y(t)$ where the covariance distance \cite{szekely2007measuring} is maximal. Our approach is somehow similar in the sense that we maximise a metric that accounts for local differences between the processes, but differs 
in two important ways: 
(i) we work under the GP assumption in order to rely on a fast and easy metric to compute (the KL divergence); and (ii) our main goal is to learn about the local maximum divergence interval. 
Some additional comments are in order: First, the discrete set $\widetilde{\mathcal{A}}^*(\tilde{c})$ does not necessarily corresponds to an interval; and second, Eq.~\eqref{optim1} entails a cumbersome combinatorial set function optimisation problem even for moderate values of $p$. Since the goal of the paper is to select a subset of the domain of the GPs instead of isolated points, we put some additional structure on the \textit{shape} of our candidate set $\mathcal{A}^*$ as in the following definition.

\begin{definition}\label{Discrete_Opt}
 Consider $\mathcal{C}_\mathcal{T}$ as the collection of all contiguous subsets from the ground set $\mathcal{T}$, such that for any  $\mathcal{A} \in \mathcal{C}_\mathcal{T}$ then $\mathcal{A} = \{t_k,t_{k+1},\dots t_{k+l}\}$ for some $1\leq k\leq p$ and $ 0 \leq l \leq p-k$. Then, the interval of local maximum KL divergence of size $c \in (0,1)$, denoted onward as $\mathcal{A}^*(c)$, is defined throughout the following set function optimisation problem:
\begin{equation}\label{optim2}
\max_{\mathcal{A} \in \mathcal{C}_\mathcal{T}} \text{KL}_\mathcal{A}(X||Y), \text{ s.t. } \text{len}(\mathcal{A}) \leq c \lambda(T),
\end{equation}
\noindent
where $\text{len}(\mathcal{A})=\max_{t\in\mathcal{A} }(t) - \min_{t\in\mathcal{A} }(t)$ is the length function corresponding to subset $\mathcal{A}$.
\end{definition}
The parameter $c \in (0,1)$ in Eq.~\eqref{optim2} determines the proportion of the domain to be selected and avoids the use of a particular length scale as is the case of $\tilde{c}$ in Eq.~\eqref{optim1}. In the empirical Section~\ref{S:Experiments} we discuss data driven strategies to choose the value of $c$. 
The existence of $\mathcal{A}^*(c)$ follows from the result stated in Proposition--\ref{prop1} point (b) and the fact that $\mathcal{P}_\mathcal{T}$ is finite (see the Appendix for further details). However, the interval of local maximum divergence $\mathcal{A}^*(c)$ does not need to be unique, as can be easily seen by considering the limiting case where GP's $X(t)$ and $Y(t)$ have the same mean and variance functions. In such a case every compact subset of $T$ with measure $c$ is a set of maximum KL local divergence. Moreover, one can learn several and disjoints intervals of maximum KL local divergence by applying sequential learning. This means selecting a new interval of maximum KL local divergence, after discarding previously selected intervals.
    
\subsection{Learning intervals with local maximum KL divergence from data}\label{ss:lern}
To learn $\mathcal{A}^*(c)$ from data we consider samples recorded over the same discrete grid $\mathcal{T}$:  $\mathcal{D}_X=\{\textbf{x}_i\}_{i=1}^n$ and $\mathcal{D}_Y=\{\textbf{y}_j\}_{j=1}^m$ drawn from $GP(\mu_X(t),\sigma_X(t,s))$ and $ GP(\mu_Y(t),\sigma_Y(t,s))$ respectively (more general sampling designs are discussed in below). Let $\boldsymbol \Psi_{\mathcal{T},X} = \{\boldsymbol\mu_{\mathcal{T},X},\boldsymbol\Sigma_{\mathcal{T},X}\}$ 
be the true parameters corresponding to GP $X(t)$, the Maximum Likelihood estimates are given by: 
$$\widehat{\boldsymbol\Psi}_{\mathcal{T},X}=\left\{\widehat{\boldsymbol\mu}_{\mathcal{T},X} =\frac{1}{n}\sum_{i=1}^n\mathbf{x}_i; \quad \widehat{\boldsymbol\Sigma}_{\mathcal{T},X}= \frac{1}{n}\sum_{i=1}^n(\mathbf{x}_i-\widehat{\boldsymbol\mu}_{\mathcal{T},X})(\mathbf{x}_i-\widehat{\boldsymbol\mu}_{\mathcal{T},X})^T\right\}, $$ 
and an analogous expression holds for $\widehat{\boldsymbol\Psi}_{\mathcal{T},Y}$ as well. 
Plugging suitable partitions $\widehat{\boldsymbol\Psi}_{\mathcal{A},X}$ and $\widehat{\boldsymbol\Psi}_{\mathcal{A},Y}$ corresponding to the subset $\mathcal{A}$ into Eq.~\eqref{optim2} gives us an estimate of $\widehat{\text{KL}}_\mathcal{A}(X||Y)$. Notice that the trace and determinant are continuous functions in the space of real symmetric matrices, then $\widehat{\text{KL}}_\mathcal{A}(X||Y)$ inherits interesting statistical properties, in particular for a fixed $p$, consistency, asymptotic normality and efficiency, providing that $p/n\to 0$, $p/m\to 0$ and $m/n\to \rho\in (0,1)$ as $n\to \infty$ and $m\to \infty$. Having in hand a consistent estimator of the local KL divergence, to learn about $\mathcal{A}^*(c)$, for $c \in (0,1)$, we follow the step in provided in Algorithm~\ref{algo}.
\begin{algorithm}\label{algo}
\caption{Estimating the interval of local maximum KL divergence from GP data.\label{alg}}   
\begin{description}
\item[Inputs:] $\mathcal{D}_X$ and $\mathcal{D}_Y$ and a length constrain $0<c\leq 1$.  
\item[~~~Step 1:] Parameter estimation. 
\begin{enumerate}
\item[IF:] $\mathcal{D}_X$ and $\mathcal{D}_Y$ are recorded over the same grid $\mathcal{T}$,  compute Maximum Likelihood estimates $\widehat{\Psi}_{\mathcal{T},X}$ and $\widehat{\Psi}_{\mathcal{T},Y}$ from data.\medskip
\item[ELSE:] Use your favourite (semi/non) parametric estimation method to compute $\widehat{\mu}_\ell(t)$ and $\widehat{\sigma}_\ell(t,s)$ from $\mathcal{D}_\ell$ for $\ell = \{X,Y\}$. Define a suitable common grid $\mathcal{T}$ and compute Maximum Likelihood estimates $\widehat{\Psi}_{\mathcal{T},X}$ and $\widehat{\Psi}_{\mathcal{T},Y}$ via the evaluation of $(\widehat{\mu}_\ell(t),\widehat{\sigma}_\ell(t,s))$ over $\mathcal{T}$ for $\ell = \{X,Y\}$ respectively.
\end{enumerate}
\item[~~~Step 2:] Exhaustive optimisation: Let $\mathcal{C}_{\mathcal{T}}(c)$ be the set of all contiguous subsets in $\mathcal{T}$ such as $\text{len}(\mathcal{A})\leq c\lambda(T)$ if $\mathcal{A} \in \mathcal{C}_{\mathcal{T}}(c)$,  then for all $\mathcal{A} \in \mathcal{C}_{\mathcal{T}}(c)$ compute $\widehat{\text{KL}}_\mathcal{A}(X||Y)$ and return: $\widehat{\mathcal{A}}^*_c \equiv \operatorname{arg\, max}_{\mathcal{A} \in \mathcal{C}_\mathcal{T}(c)}\widehat{\text{KL}}_\mathcal{A}(X||Y).$
\end{description}
\label{algo}
\end{algorithm}

Algorithm~\ref{alg} warrants some comments: Step 1 is carried only once even in the case of estimating several intervals of local maximum KL divergence for different values of $c$. 
In our \texttt{R} implementation, the set function optimisation problem in Step 2 is solved via the evaluation of $\widehat{\text{KL}}_\mathcal{A}(X||Y)$ over all sets in $\mathcal{C}_{\mathcal{T}}(c)$; nevertheless other more efficient derivative--free optimisation approaches \cite{nocedal1999numerical} can be considered as well. 
In Section~\ref{S:Experiments} we assess the computational efficiency of the estimation method entailed in Algorithm~\ref{algo} over different data generating scenarios. 

\paragraph{The curse of dimensionality} In the context of GP data, usually $n$ and $m$ are relatively small in comparison to $p=|\mathcal{T}|$, being often the case where $p\gg \max\{n,m\}$. Since the estimation of $\mathcal{A}^*(c)$ depends on the estimation of two $p\times p$ covariance matrices,  some remedy actions are necessary in order to obtain suitable estimations for $\boldsymbol \Sigma_{\mathcal{T},X}$ and $\boldsymbol \Sigma_{\mathcal{T},Y}$ from data whenever $p$ is relatively large in comparison to sample sizes $n$ and $m$. A well known strategy is to impose structure in the GP's covariance functions, assuming for instance that both processes are conditionally independent in time (i.e. $\sigma_X(t,s)=\sigma_Y(t,s)=0$ for all $t\neq s$), which corresponds to assume $\boldsymbol \Sigma_{\mathcal{T},X}$ and $\boldsymbol \Sigma_{\mathcal{T},Y}$ are diagonal covariance matrices. Another less contrived approach to circumvent the curse of dimensionality is to consider a penalised likelihood covariance matrix estimator as follows:
$$ \widehat{\boldsymbol\Sigma}_\eta = \eta\,\widehat{\boldsymbol\Sigma} + (1-\eta)\,\text{diag}(\widehat{\boldsymbol\Sigma}), \text{ for } \eta\in [0,1],$$
where $\eta$ is a regularisation parameter that shrinks the Maximum Likelihood estimator of the variance matrix towards its diagonal \cite{hastie2009elements}. The value of $\eta$ is typically determined using cross validation methods. Nevertheless, other approaches such as Banding, Tapering and alternative thresholding methods are also available, we refer to   \cite{pourahmadi2013high}[Ch.~6] and references therein for more details.

\paragraph{Sampling designs and missalignments}
In order to simplify the exposition, in section \ref{ss:KL}, we assume that $X(t)$ and $Y(t)$ are recorded over the same equally space time point grid $\mathcal{T}$, but other sampling designs are also frequent in practice. In such cases where the processes are not recorded over the same grid, to use Algorithm~\ref{alg}, we first need to define a common auxiliary grid  $\mathcal{T}$ and then proceed as follows:
\begin{enumerate}
    \item Use a suitable smoothing technique such as kernel smoothing, smoothing Splines, or GP regression among many others; and estimate the functional parameters $\mu_\ell(t)$ and $\sigma_\ell(t,s)$ for $\ell = \{X,Y\}$ from data.
    \item Consider $\widehat{\boldsymbol \mu}_\ell = (\widehat{\mu}_\ell(t_1)\dots,\widehat{\mu}_\ell(t_p))$ and $[\widehat{\boldsymbol \Sigma}]_{i,j}= \widehat{\sigma}_\ell(t_i,t_j)$, for $\ell = \{X,Y\}$ and $i=\{1,\dots,p\}$ and $j=\{1,\dots,p\}$  (i.e. the evaluation of the estimated functional parameters over the grid $\mathcal{T}$).
\end{enumerate}
After the estimation of mean vectors and covariance matrices over a common grid $\mathcal{T}$, Step~2 in Algorithm~\ref{alg} follows straightforwardly. In principle the auxiliary grid $\mathcal{T}$ contains equally spaced time points and has length $p$ according to the resolution defined by the user. Nevertheless, other sampling designs can be considered as well, for instance sampling  $\mathcal{T}$ from a multivariate \textit{prior} distribution $\pi_\mathcal{T}(t_1,\dots,t_p)$. This prior distribution encodes knowledge about the most local divergent interval in $T$ by putting more probability mass over a specific interval $A\subset T$. This procedure is a natural avenue to study a Bayesian extension of the proposed method to select domain with GP.

GP data are sometimes recorded with different types of random amplitude and time variations that produce a misalignment in data \cite{ramsay1998curve}. The GP data asynchrony may act as a confounding factor when the aim is the estimation of the interval of local maximum divergence, since the mean and variance functions estimators defined in \S~\ref{ss:lern} are useless. In this case, before the implementation of Step 1 in Algorithm~\ref{algo}, we suggest to pre-process GP data using standard alignment or synchronisation tools (a.k.a. as curve registration in functional data) such as the methods described in  \cite{berndt1994using,kazlauskaite2019gaussian} among others.

\paragraph{Smoothing} 
GP data sometimes present a low signal to noise ratio leading to problems in the estimation of the interval of local maximum divergence. To alleviate low signal to noise ratio issues, we recommend to smooth GP data using standard methods such as Natural Splines, or Kernel Regression, among others  \cite{ullah2013applications}. The smoothing process also enables us to use a finer grid of points (see \textit{sampling designs} paragraph). Once the smoothed data is obtained, they can be used as the input in Algorithm~\ref{algo}. 

\subsection{Inference and prediction}\label{ss:PI} 

\paragraph{Inference} 
The assess the variability of $\widehat{\mathcal{A}}^*(c)$ as an estimator of $\mathcal{A}^*(c)$, we resort to non-parametric bootstrap techniques. For every $c\in (0,1)$, the interval of local maximum divergence and its corresponding estimator can be parametrised in terms of a ball with centre $t_c$ and radius\footnote{Notice that $r_c$ is a nuisance parameter that only depends on $c$.} $r_c$ as follows: $ B(t_c^*,r_c)\equiv [\min_{t\in \mathcal{A}^*(c)} t,\max_{t\in \mathcal{A}^*(c)} t]$ which corresponds to $\mathcal{A}^*(c)$; and $B(\hat{t}_c^*,r_c)\equiv [\min_{t\in \widehat{\mathcal{A}}^*(c)} t,\max_{t\in \widehat{\mathcal{A}}^*(c)} t]$  which corresponds to $\widehat{\mathcal{A}}^*(c)$. Then a $1-\alpha$ bootstrap confidence interval for $t^*_c$ is given by  $\text{CI}_{1-\alpha}(t^*_c)\equiv [\hat{t}^*_{c,\alpha/2},\hat{t}^*_{c,1-\alpha/2}]$, where $\hat{t}^*_{c,\alpha}$ is the $\alpha \in (0,1)$ quantile corresponding to the empirical distribution $\widehat{F}_{B}(t^*_c)$ obtained using $B\gg 0$ bootstrap samples from GP data. From the later confidence interval, we define our $1-\alpha$ confidence set for $\mathcal{A}^*(c)$ as $\text{CS}_{1-\alpha}(\mathcal{A}^*(c))\equiv \cup_{t \in \text{CI}_{1-\alpha}(t^*_c)} B(t,r_c)$. In the experimental sections we report $\widehat{F}_{B}(t^*_c)$ and $\text{CI}_{1-\alpha}(t^*_c)$, the computation of $\text{CS}_{1-\alpha}(\mathcal{A}^*(c))$ is straightforward from the later confidence interval.\\

Some additional comments on the uncertainty quantification are in order. Regardless the experimental design, there is a trade-off between the value of $c$ and the variability of our estimator. For instance, for $p = 100$ and $c = 0.1$ there are 91 possible intervals in the domain, while for $c = 0.9$ there are only 11 intervals. Hence, the uncertainty associated to $\mathcal{\widehat{A}}^*(c)$ converges to zero as $c \to 1$--these corresponds to narrower intervals for $t^*_c$ as $c \to 1$. Other confidence interval based on bootstrap procedures can be considered as well, such as the percentile or the Student-\textit{t} method. Moreover, the parametric bootstrap is also another possible approach, taking into account the Gaussian assumptions. Notice that bootstrap confidence intervals are neither exact nor optimal, but they are largely used in practice since the method is easy to implement and its accuracy is near-exact. For a general discussion on the asymptotic properties regarding the coverage probability of a bootstrap confidence interval we refer to  \cite{efron1994introduction}.

\paragraph{Domain selection and GP classification} Domain selection is an important preliminary step before training a classification model \cite{berrendero2016variable}. In the context of GP data, the Discriminant Analysis (DA) is the Bayes Optimal Classifier  \cite{fraley2002model}; hence to classify a new unlabelled instance $\mathbf{z}$--a realisation from $X(t)$ or $Y(t)$ recorded over $\mathcal{T}$--, the DA considers the sign of the following discriminant function:
\begin{equation}\label{eq1}
\small
D_{\mathcal{T}}(\mathbf{z})=\frac{1}{2}\left((\mathbf{z}-\boldsymbol{\mu}_{\mathcal{T},Y})^T\boldsymbol{\Sigma}_{\mathcal{T},Y}^{-1}(\mathbf{z}-\boldsymbol{\mu}_{\mathcal{T},Y}) -(\mathbf{z}-\boldsymbol{\mu}_{\mathcal{T},X})^T\boldsymbol{\Sigma}_{\mathcal{T},X}^{-1}(\mathbf{z}-\boldsymbol{\mu}_{\mathcal{T},X}) - \ln\bigg(\frac{\text{det}\,\boldsymbol\Sigma_{\mathcal{T},X}}{\text{det}\,\boldsymbol\Sigma_{\mathcal{T},Y}}\Bigg)\right) +\ln\bigg(\frac{\pi_X}{\pi_Y}\bigg),
\end{equation}
where $\pi_X$ and $\pi_Y = 1 - \pi_X$ are the corresponding prior probabilities. $D_{\mathcal{T}}(\mathbf{z})$ encodes the rule to classify $\mathbf{z}$ as generated from $X(t)$ or $Y(t)$. In high dimensional contexts ($p\gg 0$), an important drawback of DA is the lack of reliable estimates of the involved mean vectors and variance matrices. This leads to inaccurate classification results in case no action is taken on the course of dimensionality. In this regard, it will be convenient to use only a small compact subset $A\subset T$ of GP data so as to compute the discrimination function. This corresponds to replacing $D_{\mathcal{T}}(\mathbf{z})$ by $D_{\mathcal{A}^*(c)}(\mathbf{z})$ for a suitable value $c\in(0,1)$, reducing the number of parameter entailed in the discrimination analysis. Choosing a suitable value for $c$ will be crucial to obtain accurate classification results; in Section~\ref{s:realdata} we show how to fix $c$ using cross validation methods to improve the discrimination power between normal and myocardial infarction heartbeats. It is interesting to mention that a choice of $c \text{ such that } \tilde{c} = c\lambda(T) = 1$, would transform the domain selection into a variable selection problem, that is, selecting a variable on the domain of the processes. While this procedure remains valid, it does not allow for the exploitation of the existing information in the covariance matrix of the processes by considering only the variability at the selected variable. 

\section{Simulation Study}\label{S:Experiments}
In this section we assess the performance of Algorithm~\ref{algo} in the estimation of $\mathcal{A}^*(c)$ via a Monte Carlo simulation study. To this end, we consider GP data $\mathcal{D}_X=\{\textbf{x}_i\}_{i=1}^n$ and $\mathcal{D}_Y=\{\textbf{y}_j\}_{j=1}^m$ in the interval $T=[0,\pi]$ recorded over a discrete grid $\mathcal{T}=\{0/p, \dots, (p - 1)/p\}$. Following the examples in Figure~\ref{fig:1}, we set 3 different scenarios.

\begin{figure}[!ht]
\centering
\begin{subfigure}[h]{0.3\textwidth}   
\caption{}
\centering 
\includegraphics[width=\textwidth]{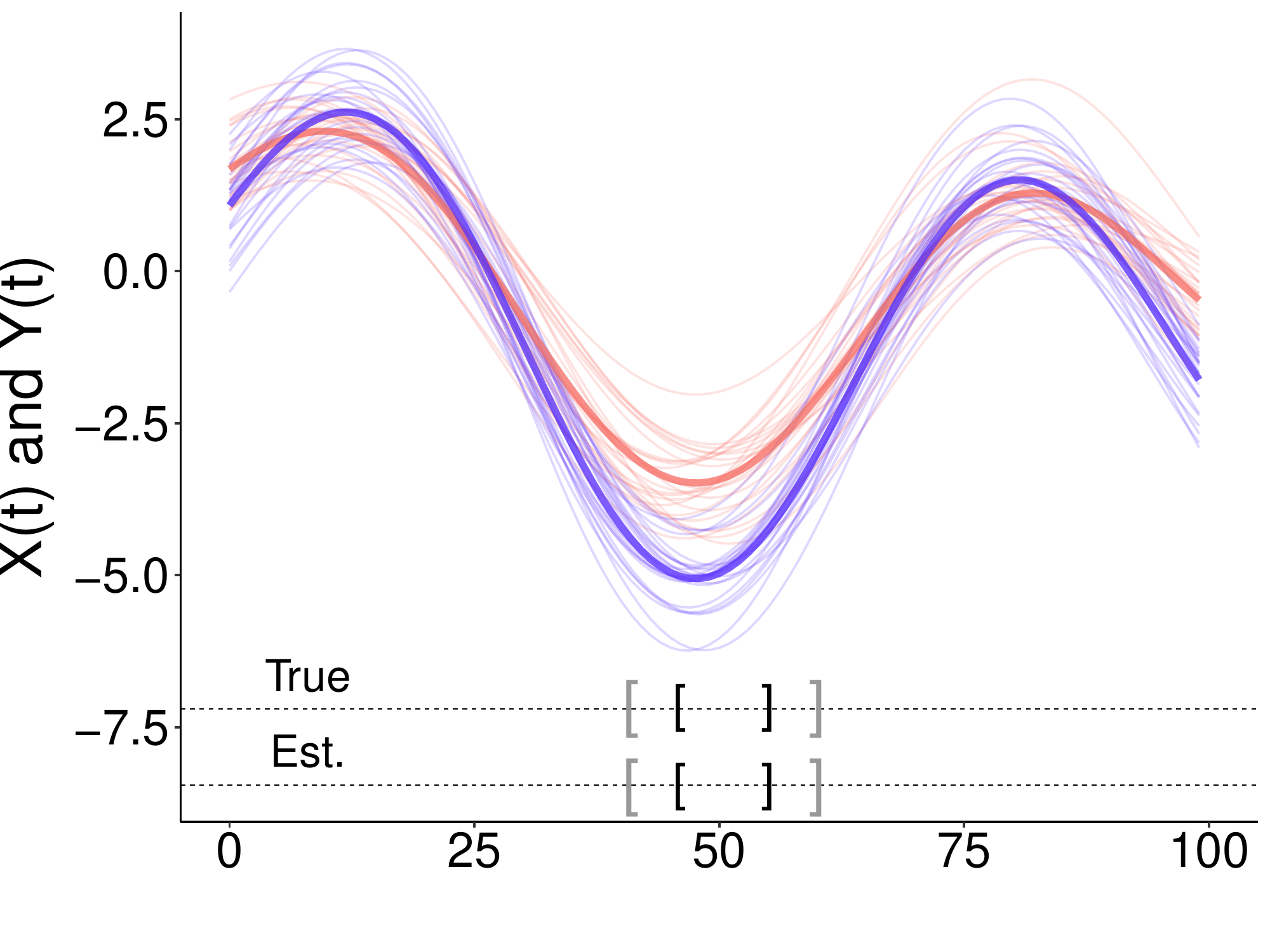}
\end{subfigure}
\begin{subfigure}[h]{0.3\textwidth}   
\caption{}
\centering 
\includegraphics[width=\textwidth]{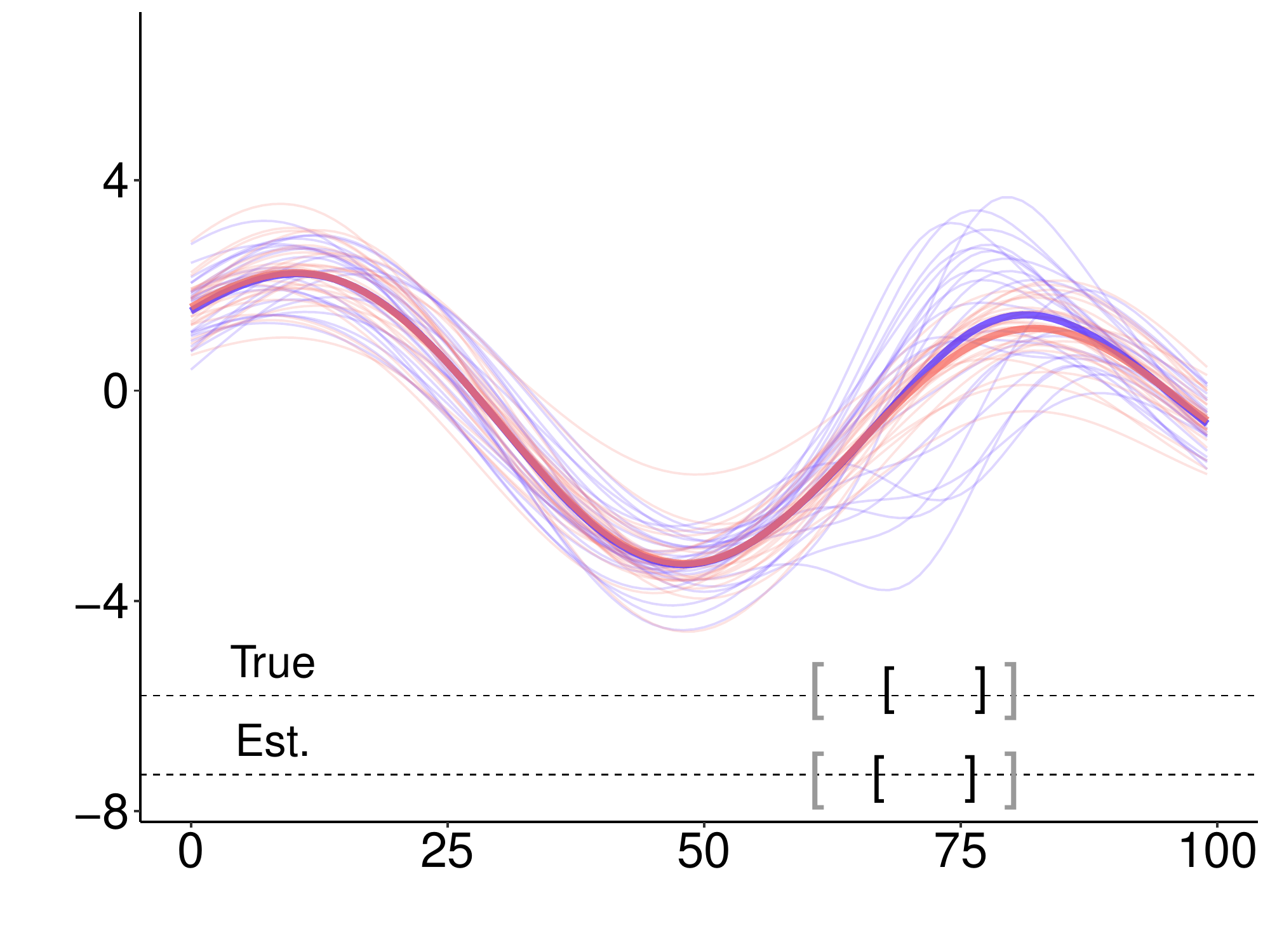}
\end{subfigure}
\begin{subfigure}[h]{0.3\textwidth}   
\caption{}
\centering 
\includegraphics[width=\textwidth]{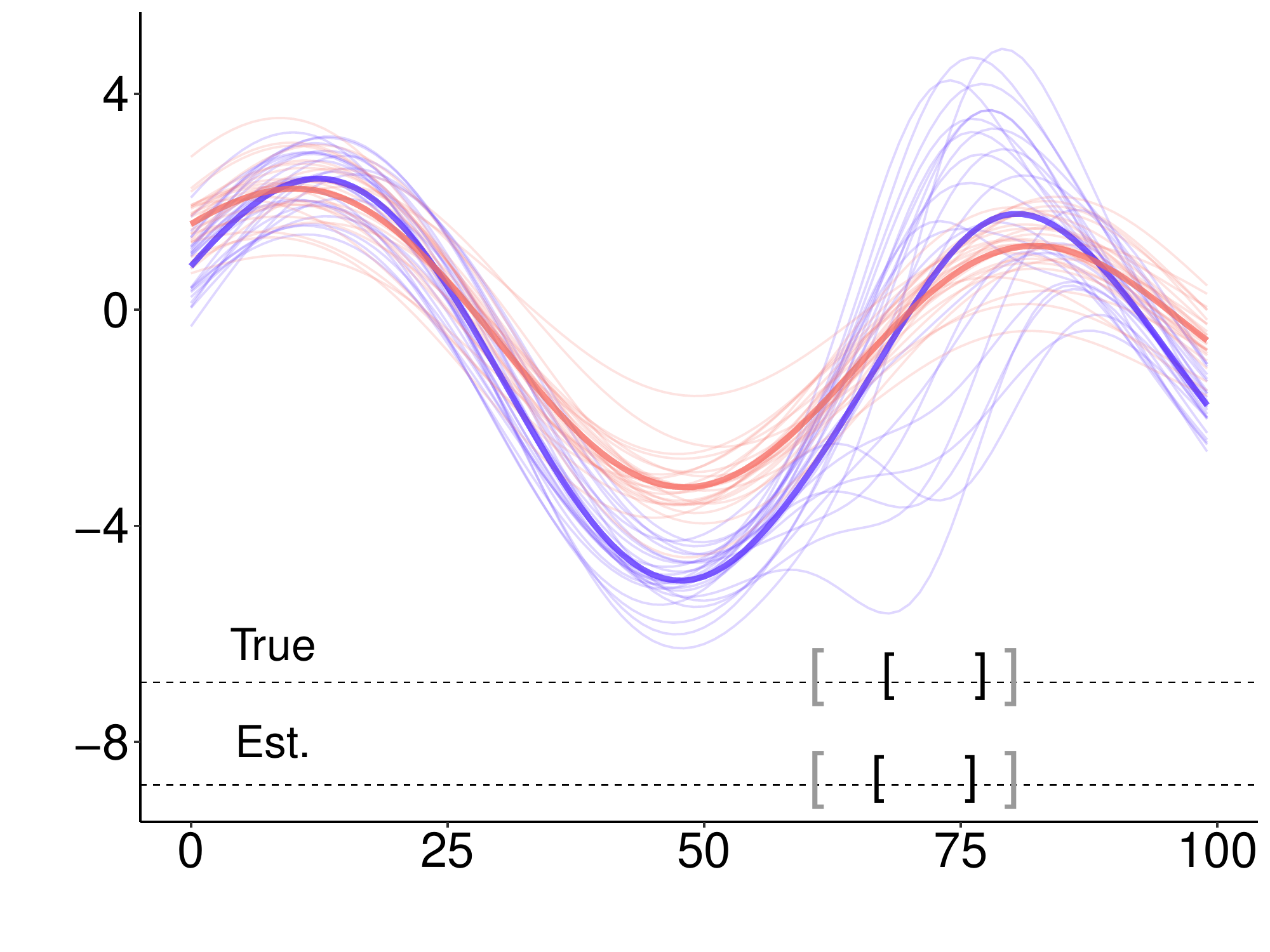}
\end{subfigure}
\begin{subfigure}[h]{0.3\textwidth}   
\centering 
\includegraphics[width=\textwidth]{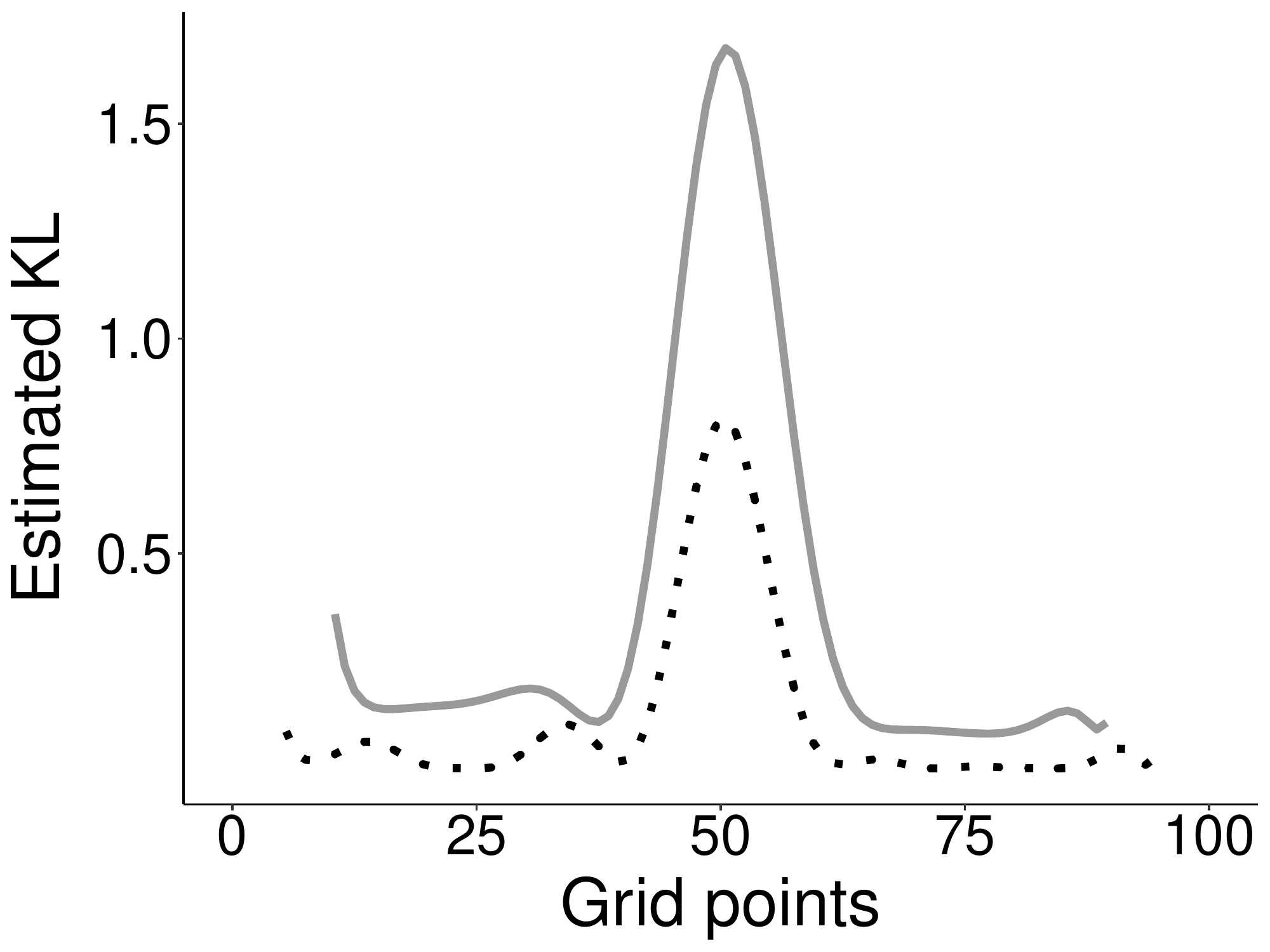}
\caption{}
\end{subfigure}
\begin{subfigure}[h]{0.3\textwidth}   
\centering 
\includegraphics[width=\textwidth]{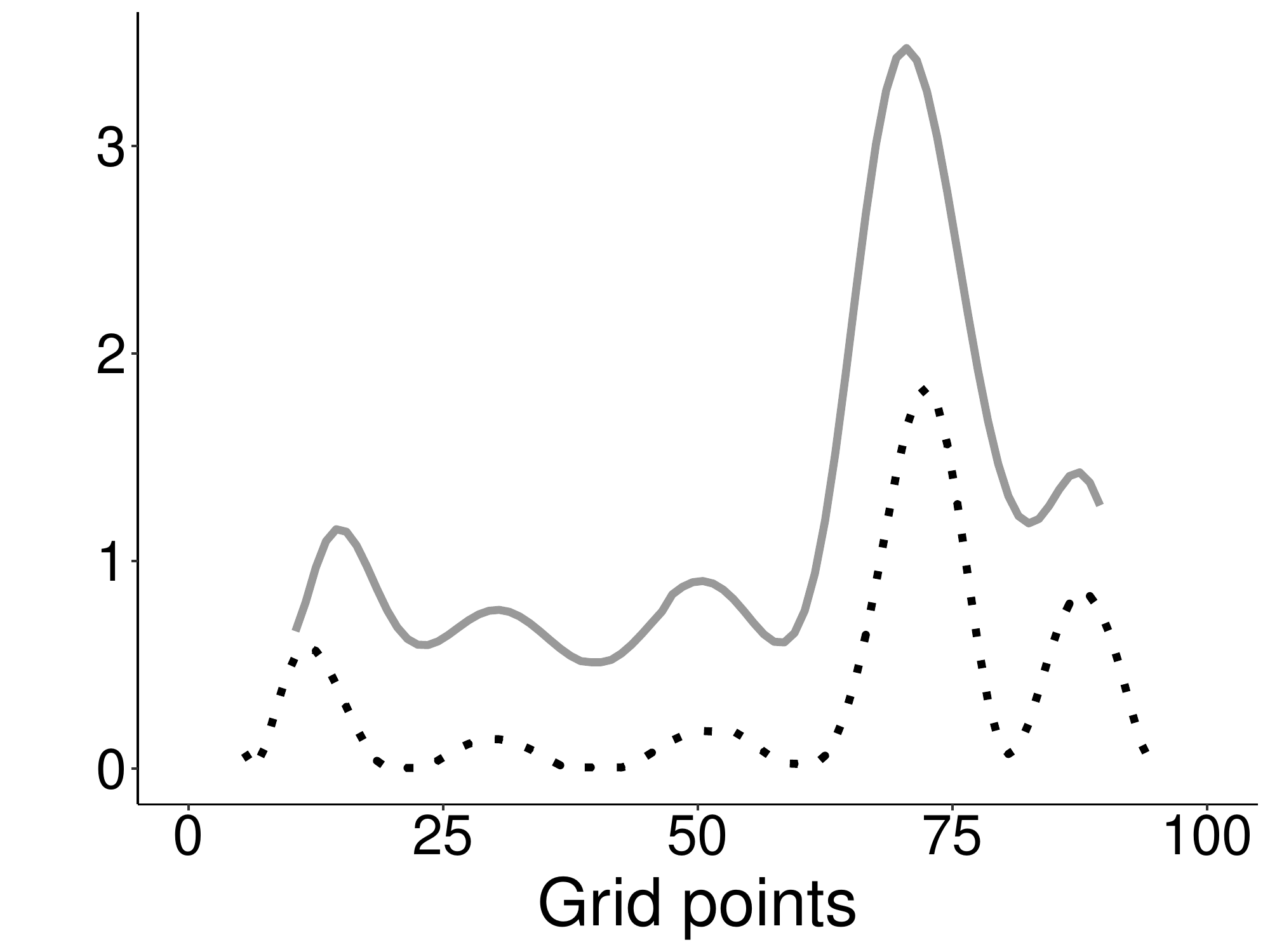}
\caption{}
\end{subfigure}
\begin{subfigure}[h]{0.3\textwidth}   
\centering 
\includegraphics[width=\textwidth]{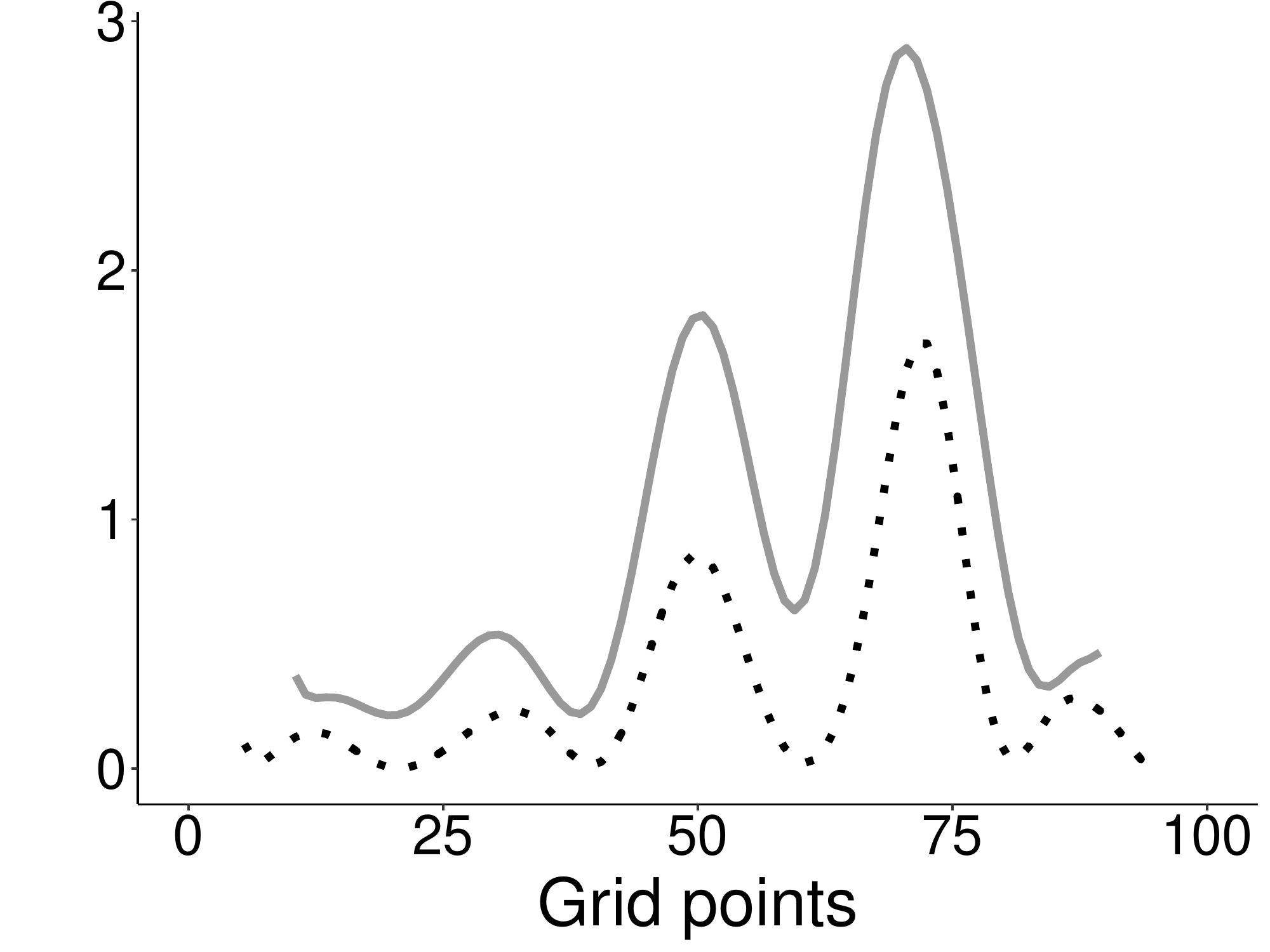}
\caption{}
\end{subfigure}
\caption{\textbf{One shot experiment:} Panels \textbf{(a)}--\textbf{(c)} illustrate the realisations of $X(t)$ and $Y(t)$ for scenarios A to C and the corresponding estimated mean functions (depicted in solid red and blue respectively). Panels \textbf{(d)}--\textbf{(f)} show the estimated local KL divergence for different interval lengths: $c = 0.1$ (\dotted)  and $c = 0.2$ (\gray{\full}); the multiple local maxima in panel (f) corresponds to the particular casuistry of scenario C (for a sufficiently large $c$, the local divergence curve will be unimodal as well).}
\label{Fig:One-Shot}
\end{figure}

\paragraph{Scenario A: Local differences in mean} 
Gaussian Process data are generated according to the following specification 
$$X(t) = (\pmb{\beta}_X + \pmb{\varepsilon})^T\pmb{\Phi}(t), \quad \text{ and } \quad 
Y(t) = (\pmb{\beta}_Y + \pmb{\varepsilon})^T\pmb{\Phi}(t),$$

\noindent
where $\pmb{\beta}_X = \{1,-2,-1,1,2,-1,2,3,-0.5\}$,   $\pmb{\beta}_Y = \{-1,-2,-1,1,2,-1,2,5,-0.5\}$, $\pmb{\Phi}(t)\equiv \{\phi_1(t),\phi_2(t),\dots,\phi_9(t)\}$ is a vector function containing the first 9 Fourier basis functions and $\pmb{\varepsilon}$ is a normally distributed random vector $ \pmb{\varepsilon}\sim \mathcal{N}_9(\textbf{0},0.25 \mathbf{I}_9)$. For the seek of simplicity, the distribution of $\pmb{\varepsilon}$ remain fixed in all scenarios. In this simulation scenario it holds that $\sigma_X(t,s) = \sigma_Y(t,s)$ for all $(t,s)\in T\times T$, nonetheless $\mu_X(t)$ is remarkably different to $\mu_Y(t)$ around grid point $50$ as can be seen in Figure~\ref{Fig:One-Shot}--(a).

\paragraph{Scenario B: Local differences in variance} Data are generated under the follow specification 
$$X(t) = \big(\pmb{\beta} + \pmb{\varepsilon}+\pmb{\gamma} e^{-(t-3\pi /4)^2}\big)^T\pmb{\Phi}(t), \quad \text{ and } \quad 
Y(t) = (\pmb{\beta} + \pmb{\varepsilon}
)^T\pmb{\Phi}(t),$$

\noindent
where $\pmb{\beta} = \{1,-2,-1,1,2,-1,2,3,-0.5\}$, and $\pmb{\gamma} \sim \mathcal{N}(\pmb{0},\tau^2\textbf{I}_{\mathbf{9}})$ is a multivariate normal random vector with $\tau^2 = 1$. Some comments on this scenario are in order: It holds that $\mu_X(t) = \mu_Y(t)$  for all $t \in T$, nevertheless $\sigma_X(t,s)$ present more differences with $\sigma_Y(t,s)$ around the grid point 75 (which corresponds to time point $t=3\pi /4$) as can be seen in Figure~\ref{Fig:One-Shot}--(b).

\paragraph{Scenario C: Local differences in mean and variance} For this scenario we consider
$$X(t) = (\pmb{\beta}_X + \pmb{\varepsilon}+\pmb{\gamma} e^{-(t-3\pi /4)^2}\big)^T\pmb{\Phi}(t), \quad \text{ and } \quad 
Y(t) = (\pmb{\beta}_Y + \pmb{\varepsilon})^T\pmb{\Phi}(t).$$

In this scenario $\pmb{\beta}_X$ and $\pmb{\beta}_Y$ take the same values as in scenario A, and $\pmb{\gamma}$ is defined likewise in Scenario B, therefore it holds that mean and variance functions differ over different subsets in $T$ as can be seen in Figure~\ref{Fig:One-Shot}--(c). \\

In Figure \ref{Fig:One-Shot} panels (a) to (c), we depict one shot GP data examples ($n=m = 25$ and $p = 100$) drawn from scenario A to C respectively; while in panels (d) to (f) we show the values of the estimated $\text{KL}_{\mathcal{A}(c)}(X||Y)$ as a function of the central point of each interval corresponding to $\mathcal{A}(c)$ for $c = \{0.10,0.20\}$ (i.e. $\text{KL}_{\mathcal{A}(c)}(X||Y) = \text{KL}_{t_c}(X||Y)$ where $t_c$ is the centre of the ball $B(t_c,r_c)=[\min_{t\in \mathcal{A}(c)} t,\max_{t\in \mathcal{A}(c)} t]$). Notice that the maximum value of the estimated local divergence in the lower panels in Figures~\ref{Fig:One-Shot} (d)--(f), corresponds to an estimated interval of local maximum divergence depicted in the upper panel of Figures~\ref{Fig:One-Shot} (a)--(c). In all cases of this one shot experiment, the estimated interval of local maximum divergence is close to the true interval of local maximum divergence. To validate the accuracy of the estimation method, we run a Monte Carlo simulation study. 

\begin{figure}[!ht]
\centering
\begin{subfigure}[h]{0.32\textwidth}  
\caption{}
\centering 
\includegraphics[width=\textwidth]{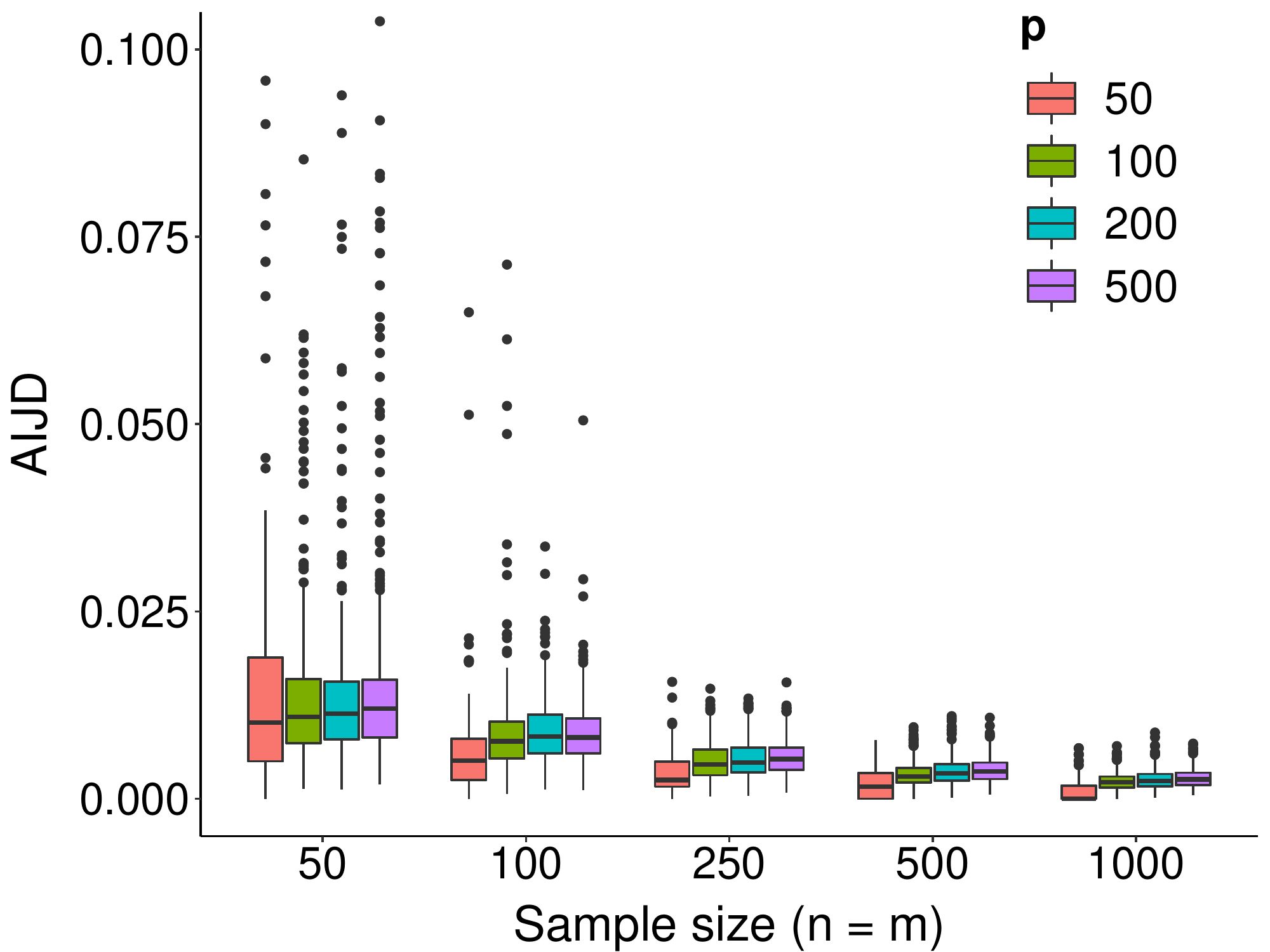}
\end{subfigure}
\begin{subfigure}[h]{0.32\textwidth}  
\caption{}
\centering 
\includegraphics[width=\textwidth]{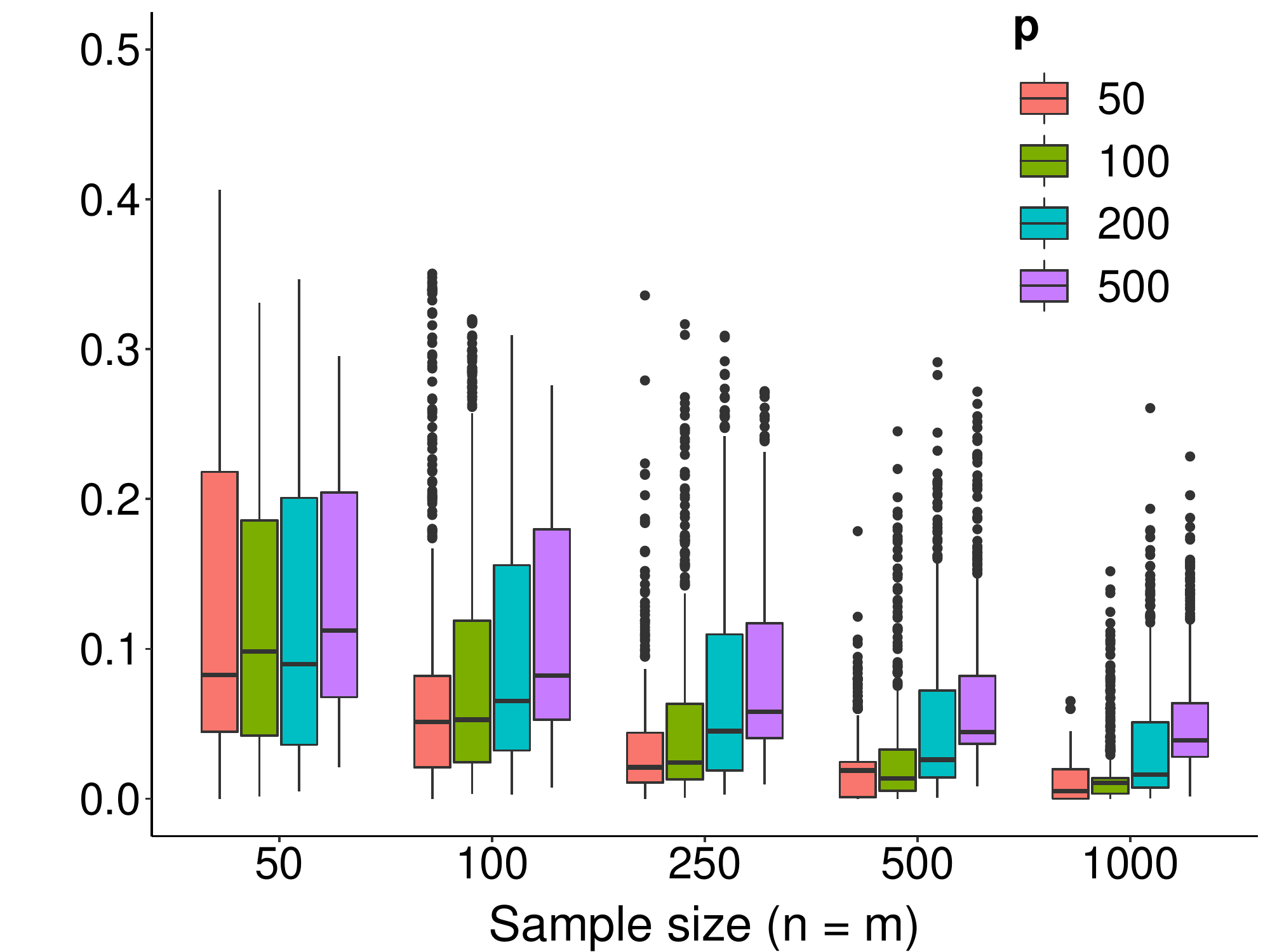}
\end{subfigure}
\begin{subfigure}[h]{0.32\textwidth}   
\caption{}
\centering 
\includegraphics[width=\textwidth]
{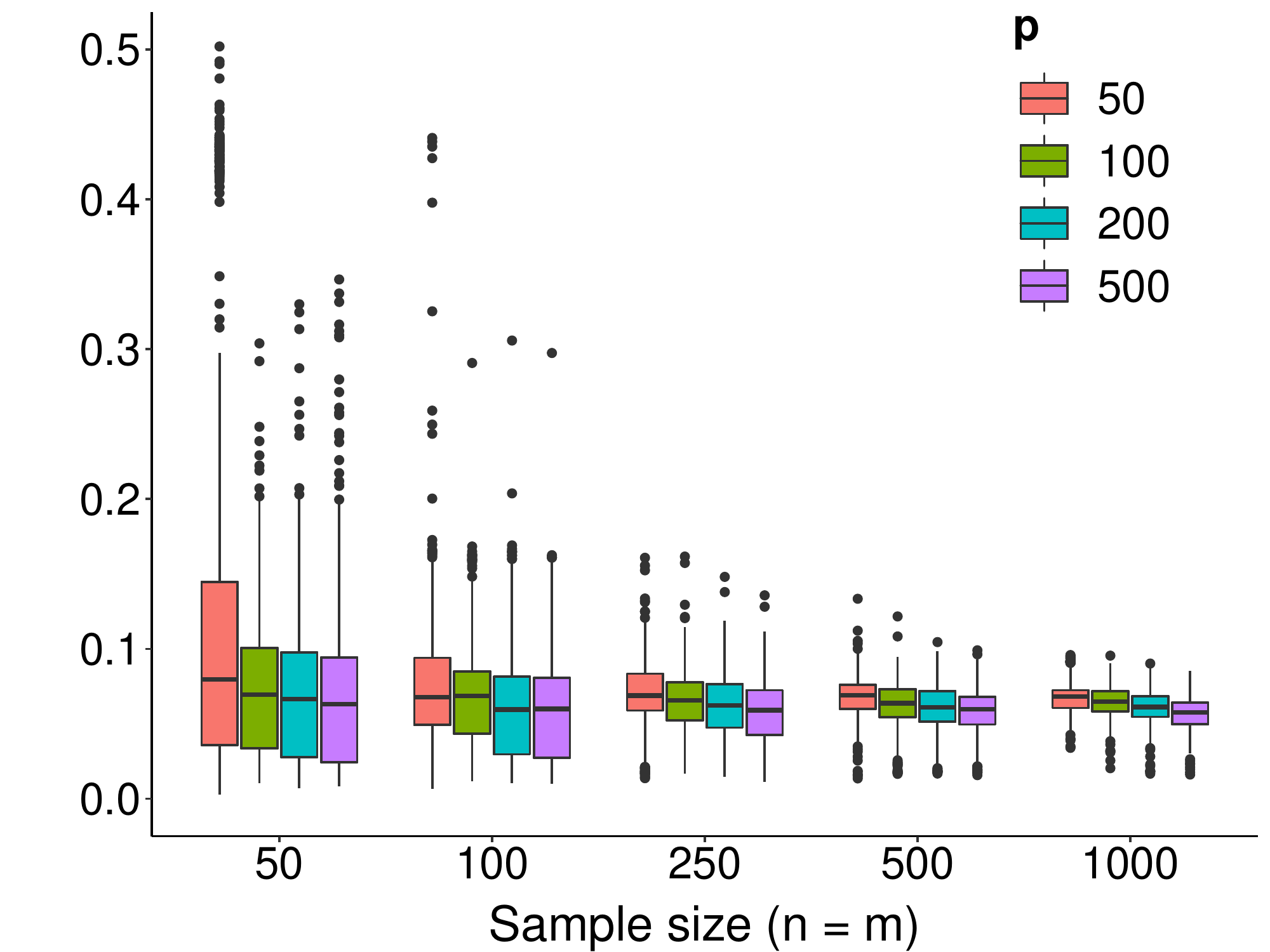}
\end{subfigure}
\caption{Empirical distribution of AIJD for different sample sizes $n$, $m$ and grid resolution levels $p$. Scenarios A--to--C in panels (a)--to--(c) respectively.
\label{Fig:MC}}
\end{figure}

\paragraph{Monte-Carlo Results}\label{ss:MonteCarlo}
The Monte Carlo simulation study consider, in each scenario, $M=1000$ data replicates for sample sizes $m=n\in \{50,100,250,500,1000\}$ and grid resolution levels $p \in \{50,100,200,500\}$. To assess the estimation performance of Algorithm~\ref{algo}, we depict in Figure~\ref{Fig:MC} the empirical distribution of the \textbf{A}verage \textbf{I}ntegrated \textbf{J}accard \textbf{D}istance (AIJD) defined as: 
\begin{equation}\label{ADJ}
  \text{AIJD} = E\Bigg\{\int_0^{1}\bigg[1 - \frac{|\mathcal{A}^*(c)\cap \widehat{\mathcal{A}}^*(c)|}{|\mathcal{A}^*(c)\cup \widehat{\mathcal{A}}^*(c)|}\bigg]\mathtt{d}c. \Bigg\}
\end{equation}
The \textit{Jaccard} Index  \cite{jaccard1912distribution} is the natural measure to assess similarity between sets, and we estimate AIJD using the trapezoidal rule over a uniform grid for $c$. The numerical analysis of our estimation method can be seen in Figure~\ref{Fig:MC}. Some comments about the Monte Carlo results are in order. As sample size $n$ and $m$ increases, then the AIJD decreases in all scenarios, this suggest a consistent estimation method. For Scenario A and C, the numerical experiment show highly accurate results even for low sample sizes (see for instance $n=m=50$), while in Scenario B, the results present more variability. Notice also that for fixed $n$ and $m$, the the estimated AIJD increases in average as $p$ increases (in almost all scenarios). This phenomena is rather natural, since as data dimension $p$ increases, the troublesome estimation of $p\times p$ covariance matrices lead to low quality local KL estimations.

\begin{figure}[!ht]
\centering
\begin{subfigure}[h]{0.49\textwidth}  \caption{}
\centering 
\includegraphics[width=\textwidth]{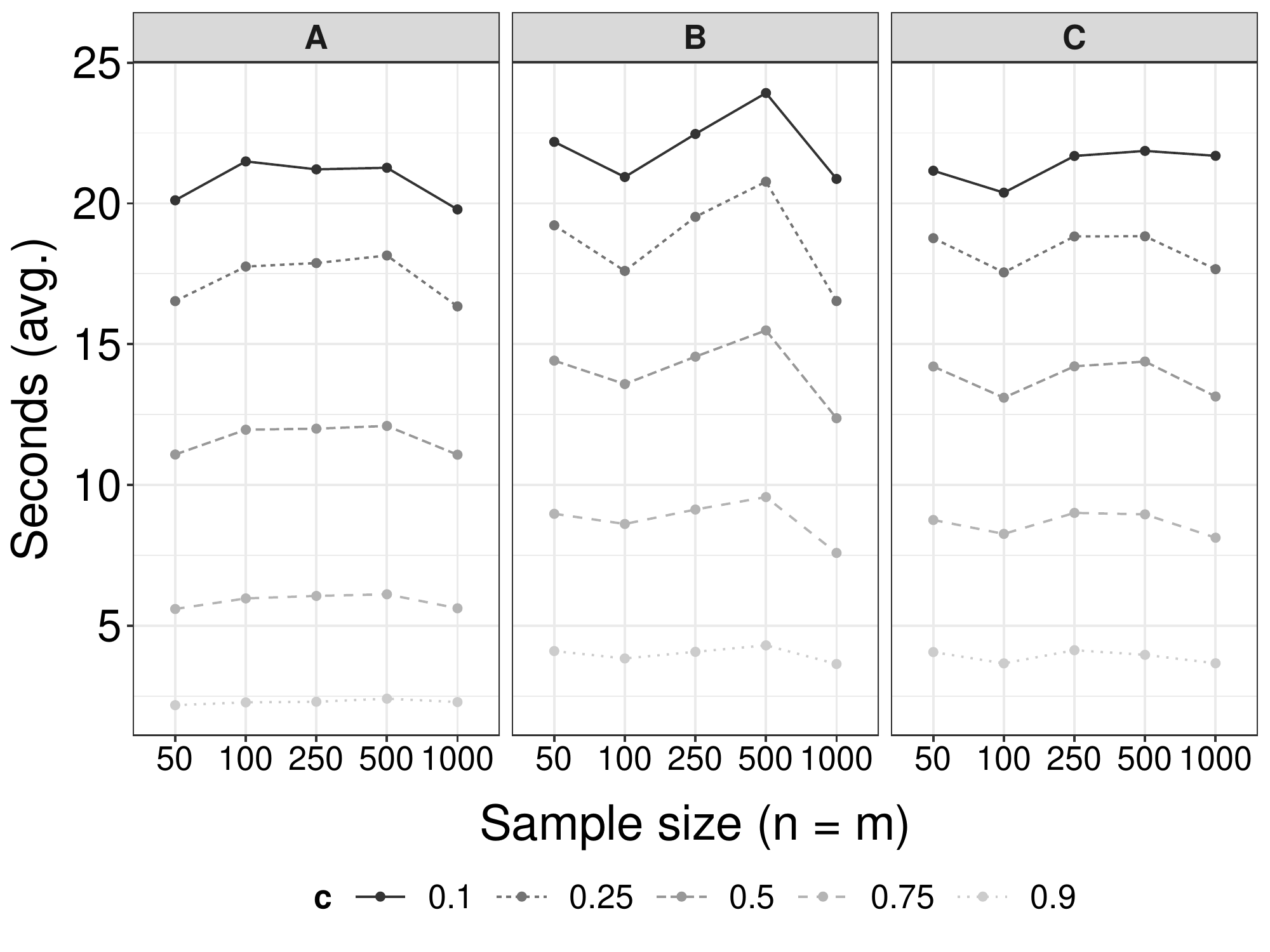}
\end{subfigure}
\begin{subfigure}[h]{0.49\textwidth}
\caption{}
\centering 
\includegraphics[width=\textwidth]{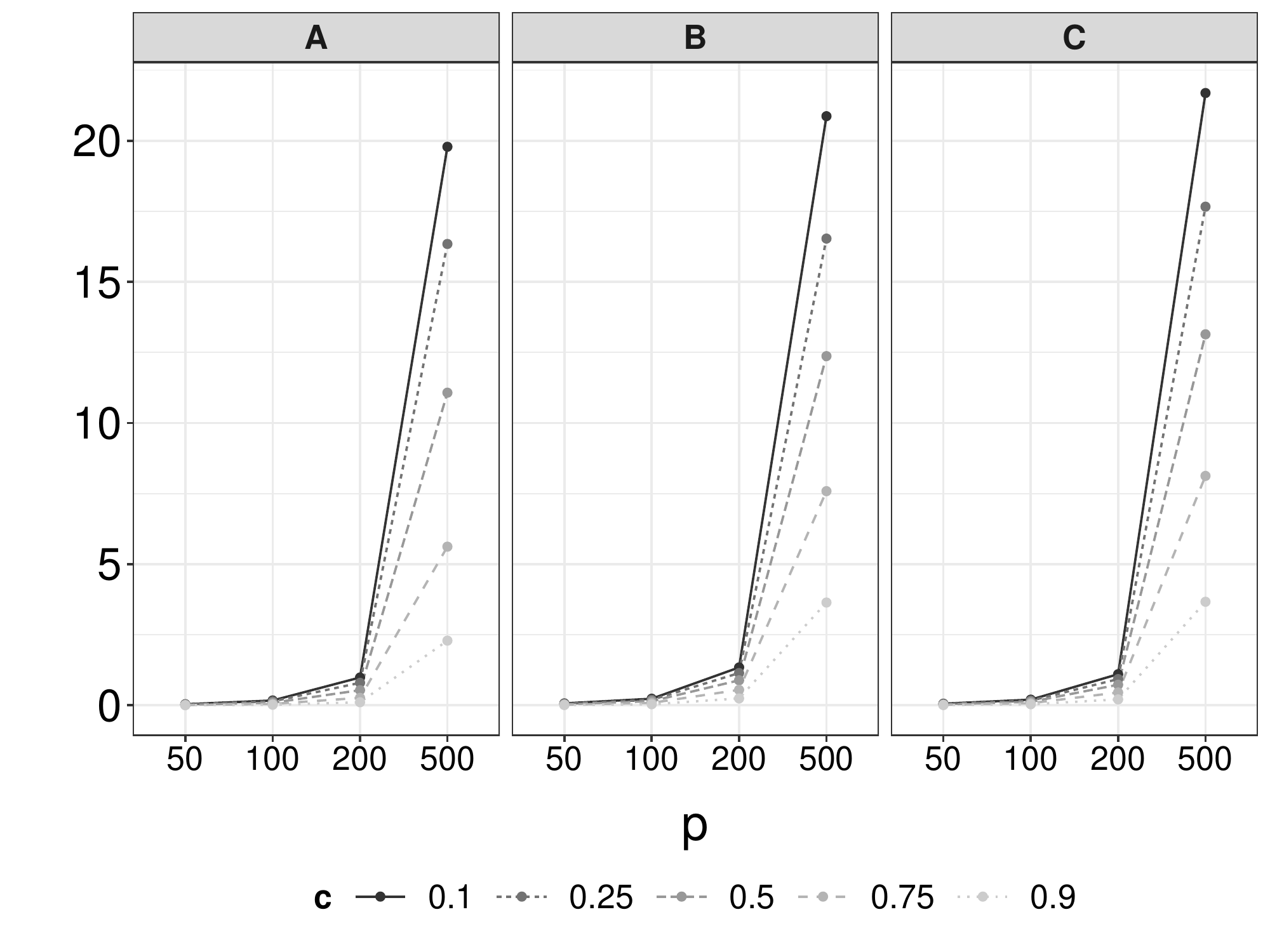}
\end{subfigure}
\caption{Average computational time required to estimate intervals of local maximum divergence for different sample sizes in \textbf{(a)} and grid resolution levels in \textbf{(b)}.}
\label{Fig:times}
\end{figure}

\paragraph{Computational complexity} 
We also study the numerical efficiency of the proposed method using the Monte Carlo simulations study\footnote{The experiment was executed in a computer with 36 cores and 1.5TB per node, paralleling per Monte Carlo iteration, sample size, grid resolution and Scenario.}. In Figure \ref{Fig:times}--(a) we depict the average time (in seconds) required to execute Algorithm~\ref{algo} for different sample sizes and fix $p=500$ (the largest value in the Monte Carlo simulation) under the 3 scenarios, while and in panel (b) we fix the sample size at its largest value $n=m = 1000$ and consider different grid resolutions levels $p$. As can be seen in both panels, small values of $c$ (short intervals) involves more computation work since Algorithm~\ref{algo} proceed in an exhaustive search for the maxima. However, even in the case of $c = 0.1$, $m=n=1000$ and $p=500$ (the shortest interval in the experiments and the largest data sets), it takes no more than 20 seconds (on average) to estimate the interval of local maximum divergence. In addition, the computational time remains constant on average as the sample size increases for fixed $p$ and $c$\footnote{Notice that the time required to compute means and variances does not increase in a sensitive way as $n$ and $m$ increases. Since mean and variances are estimated only once in Algorithm \ref{algo}, then increasing $n$ and $m$ has approximately zero impact in the computational time required to estimate local divergences and the corresponding interval of local maximum divergence.}; meanwhile for fixed $n$, $m$ and $c$, the computational time increases exponentially in $p$. Redefine Step 2 in Algorithm~\ref{algo} in order to search for a local maxima in a more efficient way will be part of the future research directions in order to tackle domain selection problems in the context of extremely large values of $p$.

\section{Monitoring electrocardiogram signals}\label{s:realdata}
The ECG signal is the visual representation of the heart electrical activity as a function of time. Learning which part of the signal spectrum is more relevant to diagnose a cardiac disease, is of fundamental importance in order to increase the probability of survival during a cardiac episode. In this section we illustrate the relevance of the proposed method in the analysis of ECG data. 

\paragraph{Data and goals}
The ECG data set \cite{olszewski2001generalized} is available in the \href{http://timeseriesclassification.com/index.php}{UEA \& UCR} repository. It consists of 200 signals sampled over a grid of 96 equally spaced instances\footnote{A healthy heart runs on a typical rate of 70 to 75 beats per minute and take about 0.8 second to complete the cardiac cycle.}. Each observation represent the cardiac electrical activity recorded during one heartbeat and there are two groups of signals: 133 normal heartbeats and 67 myocardial infarction signals. The inferential and predictive tasks relevant in this section are: i) Learn about $\mathcal{A}^*(c)$ from data and quantify the uncertainty around the estimation of such interval--we use the bootstrap strategy described in Subsection~\ref{ss:PI}. ii) Although learning about $\mathcal{A}^*(c)$ is in principle unrelated to classification, the ECG data is a popular benchmark for new classifiers; it may be sensible to ask whether the accuracy of Discriminant Analysis\footnote{DA is the optimal classifier for Gaussian data; nevertheless other classifiers can be considered as well. The analysis of other classification methods reach out of the scope of this paper.} can be improved by focusing on $\widehat{\mathcal{A}}^*(c)$ rather than treating the entire time horizon $\mathcal{T}$ equally.

\begin{figure}[!ht]
\begin{subfigure}[h]{0.5\textwidth}   
\centering \caption{}
\includegraphics[width=\textwidth]{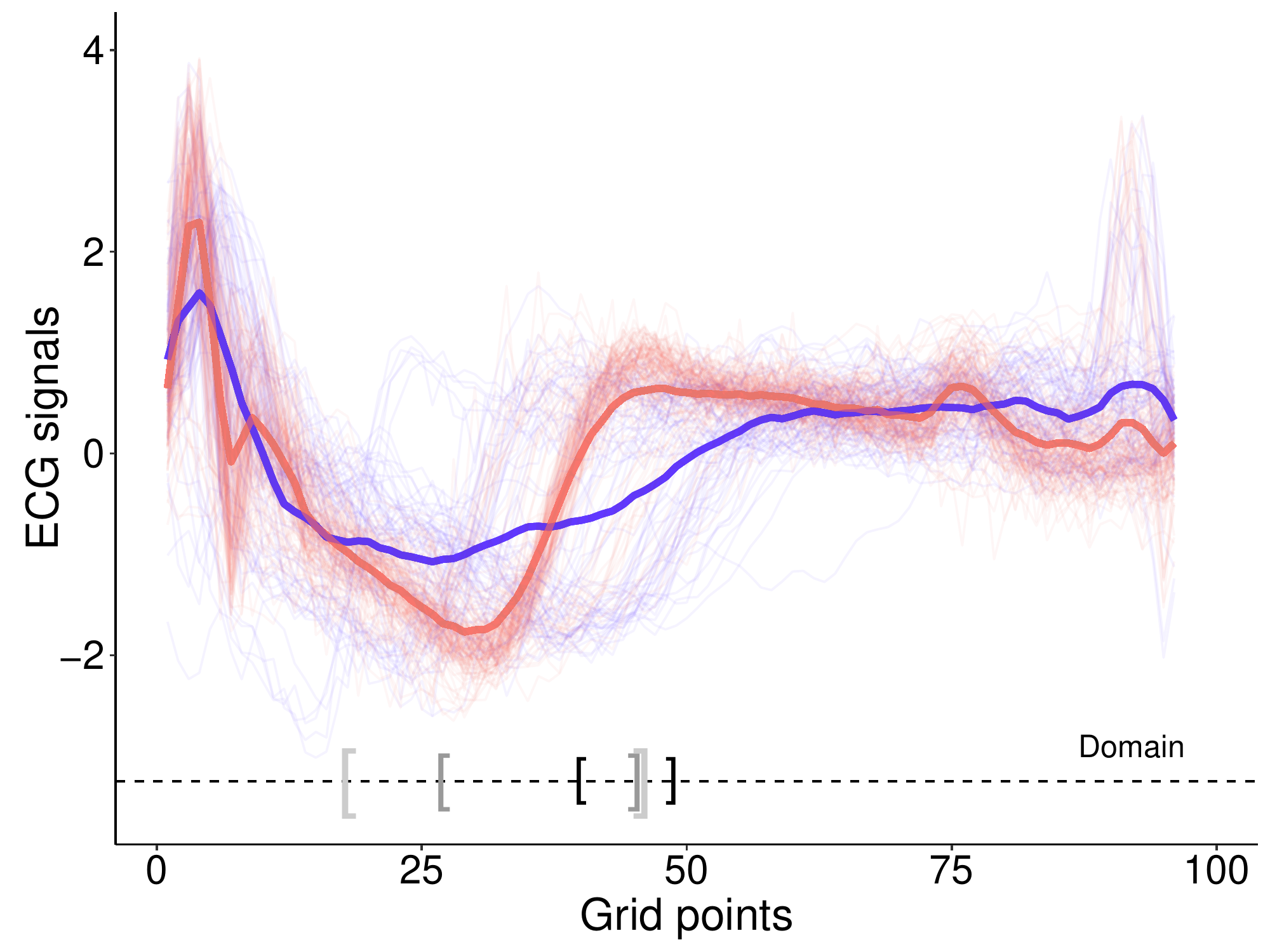}
\end{subfigure}
\begin{subfigure}[h]{0.5\textwidth}   
\caption{}
\centering 
\includegraphics[width=\textwidth]{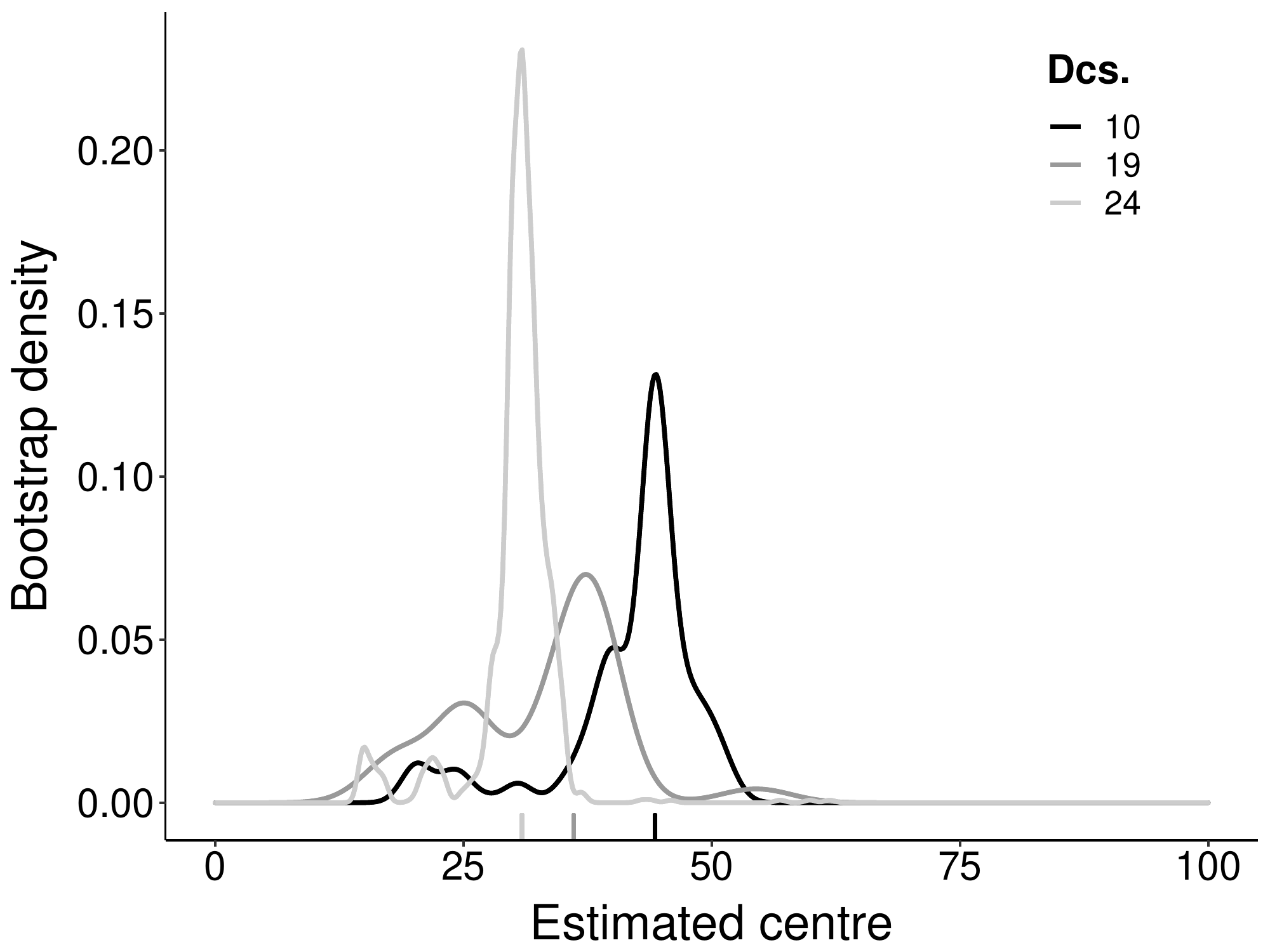}
\end{subfigure}
\caption{\textbf{(a)} ECG signals (lighted coloured curves) alongside the estimated mean functions --solid curves corresponding to healthy (\textbf{\red{---}}) and myocardial infarction (\textbf{\blue{---}}) signals. Selected domain for interval lengths $10,19$ and $24$ dcs. displayed with black "[ - ]", gray \gray{"[ - ]"} and light--gray \lightgray{"[ - ]"} brackets respectively. \textbf{(b)} 
Bootstrap densities for the interval centre (the median of each empirical distribution is reported on the horizontal axis) corresponding to intervals lengths $10,19$ and $24$ dcs.}
\label{Fig:ECG}
\end{figure}

\paragraph{Implementation and results} 
To learn about $\mathcal{A}^*(c)$, we consider $c\in\{0.1,0.2,0.25\}$ that corresponds to intervals of 10, 19 and 24 deciseconds (dcs) respectively. In Figure \ref{Fig:ECG}-(a) we display the ECG raw signals along with the corresponding estimates  $\widehat{\mathcal{A}}^*(c)$ using brackets. The selected domains corresponds to grid points between 20 and 55. All in all, for small values of $c$, the analysis suggests that while normal heartbeats and myocardial infarction signals have similar ‘peaks’ at the beginning of the sample period (i.e. they have similar Q waves, in ECG signal analysis terminology), immediately right after that period (i.e. over their ST segments) they greatly differ. To quantify the variability of our estimates, we replicate the estimation of $\widehat{\mathcal{A}}^*(c)$ using  $B=1000$ bootstrap samples. Figure~\ref{Fig:ECG}-(b) display the empirical density of the center of $\widehat{\mathcal{A}}^*(c)$ (we denote this distribution as $\widehat{F}_B(t^*)$ in \S~\ref{ss:PI}) obtained via the bootstrap samples and suggest low variability in our estimator  $\widehat{\mathcal{A}}^*(c)$ for $c\in\{0.1,0.2,0.25\}$.\\

\begin{figure}[!ht]
\begin{subfigure}[h]{0.5\textwidth}   
\caption{}
\centering 
\vspace{0.1cm}
\includegraphics[width=\textwidth]{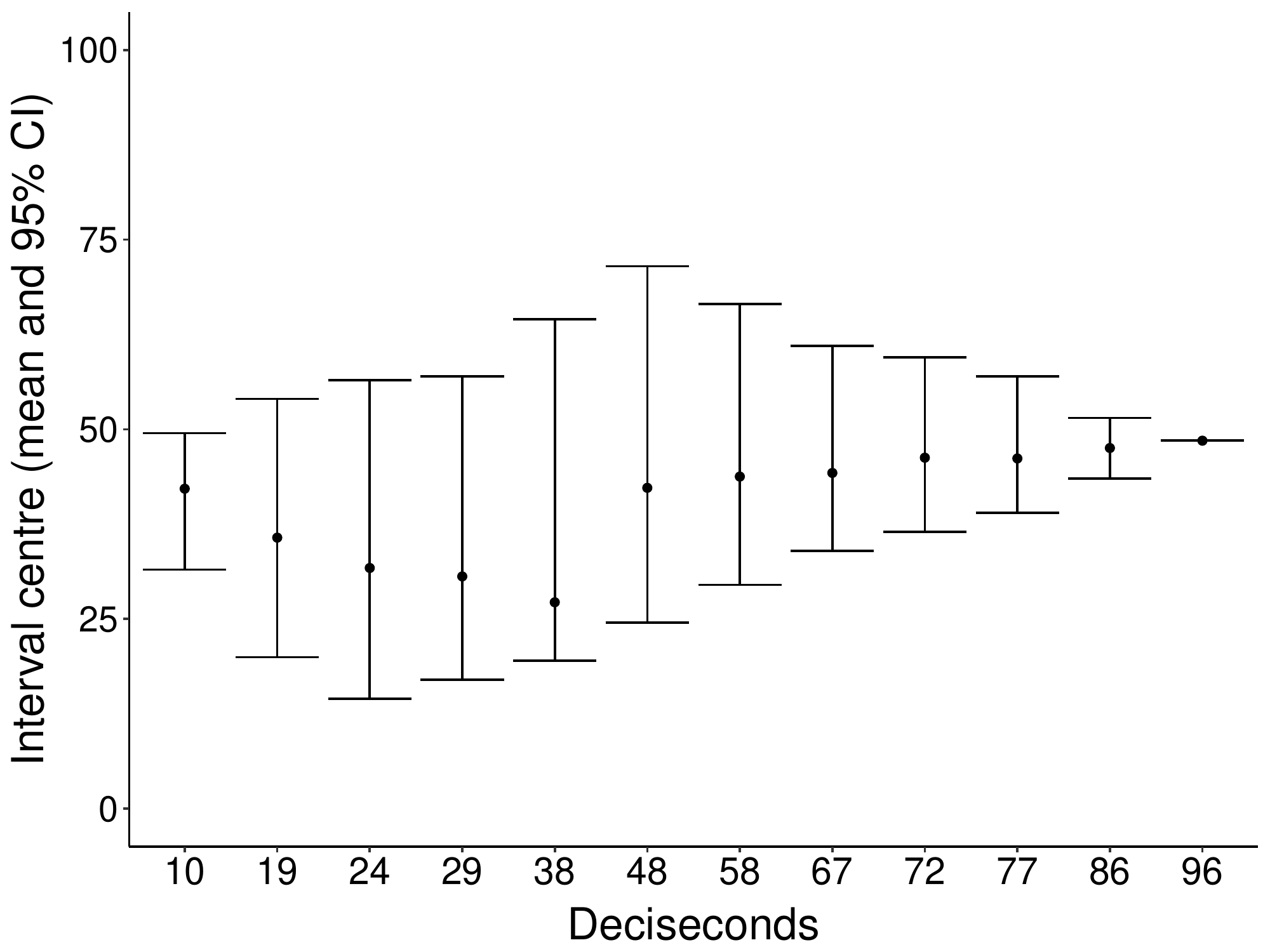}
\end{subfigure}
\begin{subfigure}[h]{0.5\textwidth}   
\caption{}
\centering 
\includegraphics[width=\textwidth]{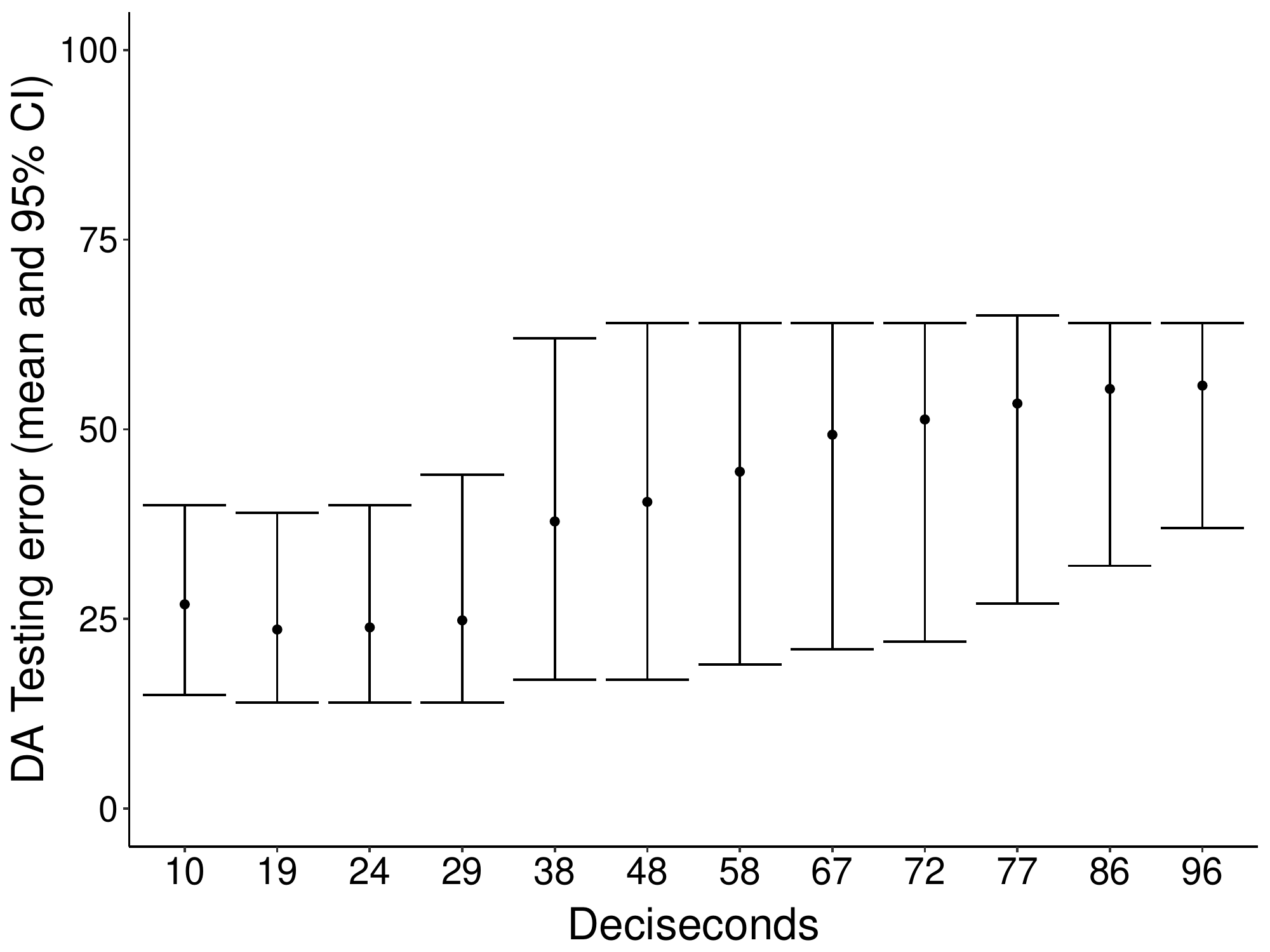}
\end{subfigure}
\caption{\textbf{(a)} Bootstrap interval centre and 95\% bootstrap CI throughout all the interval lengths considered. \textbf{(b)} Classification error using a Discriminant model for different selected domains.}
\label{Fig:ECG2}
\end{figure}

To assess the classification performance of Discriminant Analysis over different slices on the domain, we consider $c \in \{0.1,0.2,0.25,0.3,0.4,0.5,0.6,0.7,0.75,0.8,0.9,1.0\}$ which corresponds to intervals of  
$10, 19, 24, 29, 38, 48, 58, 67, 72, 77, 86$, and $96$ deciseconds respectively; and randomly split the data into training--testing samples in a 50\%-50\% fashion. For each value of $c$, we learn $\widehat{\mathcal{A}}^*(c)$ and the corresponding parameters of the discriminant function with train data; while test data is used to estimate the missclassification error rate $\widehat{\text{err}}(c)$. To assess the estimation variability on $\widehat{\mathcal{A}}^*(c)$ and $\widehat{\text{err}}(c)$, we consider $B=1000$ bootstrap samples with train and test data respectively. In Figure~\ref{Fig:ECG2}-(a) we show the estimated centres of $\widehat{\mathcal{A}}^*(c)$ (black dots in the vertical box--plots corresponds to $\hat{t}^*_c$) and the corresponding 95\% bootstrap confidence interval for such centres as a function of $c$ (in a deciseconds scale). In Figure~\ref{Fig:ECG2}-(b) we display the estimated $\widehat{\text{err}}(c)$ (black dots in the vertical box--plots) and its corresponding  95\% bootstrap confidence interval as a function of $c$ (in a deciseconds scale). As can be seen from Figure~\ref{Fig:ECG}-(b), the discrimination power of DA is significantly larger if we consider a small interval of local maximum divergence on the ECG data--say $c\in\{0.1,0.2,0.25,0.3\}$-- rather than considering the full domain corresponding to ECG data.

\section{Discussion and closing remarks}\label{S:discusion}
From a methodological outlook, the main goal of this article entails the combination of Kullback--Leibler divergence and Gaussian Processes to develop a super-fast and easy to implement algorithm for \textit{domain selection}. We define a local KL divergence, introduce its fundamental properties, and prove the existence of an interval of local maximum divergence $\mathcal{A}^*(c)$ under suitable conditions on the mean and variance functions. In addition, we also propose an estimator for $\mathcal{A}^*(c)$, devise a nonparametric approach to assess the estimation uncertainty, and also discuss relevant variants and extensions.\\

Through a Monte Carlo simulation study we numerically assess the consistency of our estimator and demonstrate that even for large $n$ and $p$ (i.e. $n=m = 1000$ and $p=500$, which corresponds to high--dimensional and large data sets), Algorithm~\ref{algo} is very efficient. Learning about intervals of local maximum divergence in the context electrocardiogram data contributes to improve diagnostic tools, as we demonstrate in the analysis of ECG data. In addition, we also explore how the discrimination power of DA (healthy vs disease heartbeats) can be improved by making emphasis on a small interval of ECG data rather than using the full domain.\\

Although the proposed method can be used for domain selection with GP, there are natural opportunities for further analysis: (i) In this paper consistency is illustrated whit a Monte Carlo simulation. Further theoretical developments will be conducted in order to establish general conditions in the model in order to ensure that $\widehat{\mathcal{A}}^*(c)$ is a consistent estimator for $\mathcal{A}^*(c)$. (ii) To define local KL divergences for GP we rely on a discretized version of $X(t)$ and $Y(t)$, therefore a natural point to address is to extend Eqs.~\eqref{Eq:KL} and \eqref{Eq:KL_local} to the non-discretized case; studying also the impact on the numerical complexity of adapting Algorithm~\ref{algo} to such context. (iii) Another important avenue for future research is the extension of the proposed domain selection method to address non GP data, perhaps using alternative metrics to assess differences between non Gaussian processes. (iv) In the framework of curve classification, is important to balance predictive classification accuracy against the number of covariates which, in the case of curves,  corresponds to the length of the interval. Therefore a natural follow-up within the remit of this paper, is the analysis of domain selection as a shrinkage method for curve classification. (v) We also propose to study alternative ways to introduce Bayesian tools for domain selection with GP as we discuss in the paragraph \textit{Sampling Designs} in \S~\ref{ss:lern}. (vi) Finally, from a computational view point, Step 2 in Algorithm~\ref{algo} needs to be reformulated for extremely high--dimensional data since its computational complexity grows exponentially in $p$ as we  mention in Section~\ref{DomainSelection} and illustrate numerically in Section~\ref{ss:MonteCarlo}.

\section*{Declarations}\footnotesize 


\noindent \textbf{Conflicts of interest}: None of the authors has a conflict of interest. \vspace{0.2cm}

\noindent \textbf{Ethics approval}: Authors have no affiliations with or involvement in any organisation or entity with any financial interest or non-financial interest in the subject matter or materials discussed in this manuscript. \vspace{0.2cm}


\noindent \textbf{Consent for publication}: Authors give consent for publication. \vspace{0.2cm}

\noindent \textbf{Availability of data and code}: Data is available on \href{http://timeseriesclassification.com/index.php}{UEA \& UCR} and source code to reproduce the results is included as a supplementary file in the submission. \vspace{0.2cm}

\noindent \textbf{Authors’ contributions}: Authors contributed equally to this work.

\noindent

\appendix
\begin{myproof}[Proof of Proposition~\ref{prop1}]
\noindent
Property (a) follows directly from the properties of KL divergence. To prove the upper bound in (b), let $\boldsymbol\Sigma_{\mathcal{T},X}$ and $\boldsymbol\Sigma_{\mathcal{T},Y}$ be positive definite (PD) matrices, then $\boldsymbol \Gamma_\mathcal{T} \equiv \boldsymbol \Sigma_{\mathcal{T},Y}^{-1}\boldsymbol\Sigma_{\mathcal{T},X}$ is also PD with eigenvalues $\infty > \gamma_{\mathcal{T},1} \geq \dots \geq \gamma_{\mathcal{T},p} >0 $  where $p = |\mathcal{T}|$; then if  $\|\Delta_\mathcal{T}\|_2 < \infty$ it holds from Eq.~\eqref{Eq:KL} that:
\begin{equation*}
2\text{KL}_\mathcal{T}(X||Y) =  \sum_{i=1}^p \big\{\gamma_{\mathcal{T},i} - \ln(\gamma_{\mathcal{T},i}) \big\}- p + \Delta_\mathcal{T}^{\mathsf{T}}\boldsymbol\Sigma_{\mathcal{T},Y}^{-1}\Delta_\mathcal{T} < \infty,
\end{equation*}
\noindent
since $ 1 \leq x-\ln(x)< \infty$, for all $0<x<\infty$. Moreover,  $\text{KL}_\mathcal{A}\equiv \text{KL}_\mathcal{A}(X||Y) \leq \text{KL}_{\mathcal{A}^{\prime}}$ for all $\mathcal{A}^{\prime} \subseteq \mathcal{A} \subseteq \mathcal{T}$ with $|\mathcal{A}^{\prime}|\equiv d^{\prime}\leq |\mathcal{A}| \equiv d$ if the eigenvalues of  $\boldsymbol \Gamma_{\mathcal{A}^{\prime}}$ are bigger than 1. To prove the assertion, consider:
\begin{equation*}
\begin{split}
2(\text{KL}_\mathcal{A} - \text{KL}_{\mathcal{A}^{\prime}}) =  \underbrace{\sum_{i=1}^{d^\prime} \big\{\gamma_{\mathcal{A},i} - \gamma_{\mathcal{A}^{\prime},i} -\big(\ln(\gamma_{\mathcal{A},i})-\ln(\gamma_{\mathcal{A}^{\prime},i})\big) \big\}}_{A} + \\ \underbrace{\sum_{j=d^{\prime}+1}^{d} \big\{\gamma_{\mathcal{A},j} -\ln(\gamma_{\mathcal{A},j}) \big\} - (d-d^\prime)}_{B} + \underbrace{\big(\Delta_\mathcal{A}^{\mathsf{T}}\boldsymbol\Sigma_{\mathcal{A},Y}^{-1}\Delta_\mathcal{A} - \Delta_{\mathcal{A}^{\prime}} ^{\mathsf{T}}\boldsymbol\Sigma_{\mathcal{A}^{\prime},Y}^{-1}\Delta_{\mathcal{A}^{\prime}}  \big)}_{C }
\end{split}
\end{equation*}\medskip
\noindent
then $C\geq 0$ since $\Delta_{\mathcal{A}^{\prime}} ^{\mathsf{T}}\boldsymbol\Sigma_{\mathcal{A}^{\prime},Y}^{-1}\Delta_{\mathcal{A}^{\prime}} \leq \Delta_\mathcal{A}^{\mathsf{T}}\boldsymbol\Sigma_{\mathcal{A},Y}^{-1}\Delta_\mathcal{A}$ for all   $\mathcal{A}^{\prime} \subseteq \mathcal{A}$; $B\geq 0$ since  $ x-\ln(x)\geq 1$ if $0<x<\infty$; and $A\geq 0$ by the eigenvalues interlacing inequality (i.e. $\gamma_{\mathcal{A},1} \geq \gamma_{\mathcal{A}^{\prime},1} \geq \gamma_{\mathcal{A},2} \geq \gamma_{\mathcal{A}^{\prime},2}   \geq \dots \gamma_{\mathcal{A},d'} \geq \gamma_{\mathcal{A'},d'}>0$), and Napier's inequality (i.e. $\ln(x) - \ln(y)\leq (x-y)/y$ for $x\geq y > 0$):
\begin{equation*}
A = \sum_{i=1}^{d^\prime} \big\{\gamma_{\mathcal{A},i} - \gamma_{\mathcal{A}^{\prime},i} -\big(\ln(\gamma_{\mathcal{A},i})-\ln(\gamma_{\mathcal{A}^{\prime},i})\big) \big\} \geq \sum_{i=1}^{d^\prime} \frac{(\gamma_{\mathcal{A},i}-\gamma_{\mathcal{A}^{\prime},i})(\gamma_{\mathcal{A}^{\prime},i}-1)}{\gamma_{\mathcal{A}^{\prime},i}} \geq 0,    
\end{equation*}
if $\infty>\gamma_{\mathcal{A}^{\prime},1}\geq \dots \geq \gamma_{\mathcal{A}^{\prime},d^{\prime}} \geq 1$. 
Notice that $\mathcal{A}^*(c)$, the solution of the set function optimisation problem stated in Eq.~\eqref{optim2}, exist under the previous stated conditions on the mean and variance functions,  since $\mathcal{P}_\mathcal{T}$ is a finite collection of sets and $\text{KL}_\mathcal{A}(X||Y)$ is a non-decreasing and bounded set function.
\noindent
To prove property (c), consider with out loss of generality the following sequence of non--decreasing sets $\mathcal{A}^{(k)}\equiv \{t_1,t_2,\dots,t_k\}$ and notice that $\cup_{k=1}^\infty\mathcal{A}^{(k)} = \cup_{k=1}^p\mathcal{A}^{(k)} = \mathcal{T} $, then the left--continuity follows from:  
$$\lim_{k\to \infty } 2\text{KL}_{\mathcal{A}^{(k)}} = \lim_{k\to p } 2\text{KL}_{\mathcal{A}^{(k)}} = \lim_{k\to p } \sum_{i=1}^k \big\{\gamma_{\mathcal{A}^{(k)},i} - \ln(\gamma_{\mathcal{A}^{(k)},i}) \big\}- k + \Delta_{\mathcal{A}^{(k)}}^{\mathsf{T}}\boldsymbol\Sigma_{\mathcal{A}^{(k)},Y}^{-1}\Delta_{\mathcal{A}^{(k)}}=2\text{KL}_{\mathcal{T}},$$
since $\gamma_{\mathcal{A}^{(k)},i} = \gamma_{\mathcal{T},i}$  for all $i\in \{1,\dots,p\}$ and  $\Delta_{\mathcal{A}^{(k)}}^{\mathsf{T}}\boldsymbol\Sigma_{\mathcal{A}^{(k)},Y}^{-1}\Delta_{\mathcal{A}^{(k)}} = \Delta_{\mathcal{T}}^{\mathsf{T}}\boldsymbol\Sigma_{\mathcal{T},Y}^{-1}\Delta_{\mathcal{T}}$ if $k=p$. 
To prove the  right--continuity consider, without loss of generality, the sequence of non--increasing sets $\mathcal{A}_{(k)}\equiv \{t_k,t_{k+1},\dots,t_p\}$ and then $\cap_{k=1}^\infty\mathcal{A}_{(k)} = \cap_{k=1}^p\mathcal{A}_{(k)} = \{t_p\} $. The rest of the proof is tantamount to the previous case.
\end{myproof}

\bibliographystyle{unsrtnat}
\bibliography{references}  






\end{document}